\documentclass[%
 reprint,
superscriptaddress,
nofootinbib,
 amsmath,amssymb,
 aps,
pra,
]{revtex4-2}

\makeatletter
\renewcommand\@appendixcntformat[1]{\csname the#1\endcsname}
\makeatother

\usepackage[colorlinks=true,citecolor=Cerulean,linkcolor=RubineRed,urlcolor=Cerulean]{hyperref}
\usepackage{graphicx}
\usepackage{dcolumn}
\usepackage{bm}
\usepackage{appendix}
\usepackage{algpseudocode}
\usepackage{algorithm2e}
\usepackage{physics}
\usepackage{mathrsfs}

\newcommand{\mr}[1]{\mathrm{#1}}
\newcommand{\ha}[1]{\hat{#1}}

\usepackage[normalem]{ulem}

\usepackage{color}
\usepackage[usenames,dvipsnames]{xcolor}
\usepackage[colorlinks=true,citecolor=Cerulean,linkcolor=RubineRed,urlcolor=Cerulean]{hyperref}

\usepackage{cancel}

\newcommand{\eq}{\textrm{eq}}
\newcommand{\dg}{^\dagger}

\begin{document}

\preprint{APS/123-QED}

\title{Förster resonance energy transfer with transient coherent effects}



\author{Maximilian Meyer-Mölleringhof}
\thanks{These authors contributed equally to this work}
\affiliation{Department of Physics and Materials Science, University of Luxembourg, L-1511 Luxembourg}
\author{Pablo Martinez-Azcona}
\thanks{These authors contributed equally to this work}
\affiliation{Department of Physics and Materials Science, University of Luxembourg, L-1511 Luxembourg}
\affiliation{%
 Faculty of Mathematics and Physics, Charles University, Ke Karlovu 5, CZ-121 16 Prague 2, Czech Republic
}
\affiliation{Currently: Walther-Meißner-Institut/Technical University of Munich/MCQST, 85748 Garching, Germany}

\author{Aur\'elia Chenu}
\affiliation{Department of Physics and Materials Science, University of Luxembourg, L-1511 Luxembourg}
\author{Tom\'{a}\v{s} Man\v{c}al}
\email{tomas.mancal@matfyz.cuni.cz}
\affiliation{%
 Faculty of Mathematics and Physics, Charles University, Ke Karlovu 5, CZ-121 16 Prague 2, Czech Republic
}
%

\date{\today}

\begin{abstract}
We formulate the weak intramolecular coupling Förster resonance energy transfer theory in a form suitable for calculating ultrafast non-linear response of molecular systems. We introduce a formally exact time-dependent factorization of  the molecular statistical operator into the system and bath components. Combining this factorization with unperturbed environment evolution, we generalize the traditional Förster master equation for the state population probabilities into a complete master equation for the system's reduced statistical operator. The traditional Förster theory applies in the limit where the intermolecular coupling is weak and the system-bath coupling is strong.  Our technique of derivation explicitly leads to a time non-local Förster type master equation which remains valid also in the limit of vanishing system-bath coupling. The theory predicts a rapid initial coherent evolution of populations arising from a transient initial coherence-dependent term, which induces a ‘slippage’ of the initial condition that persists during subsequent rate-controlled transfer. Comparison with exact numerical results confirms the clear improvement of the present generalization over earlier formulations of the Förster theory and delineates its range of validity.
\begin{center}
\includegraphics[width=.5\linewidth]{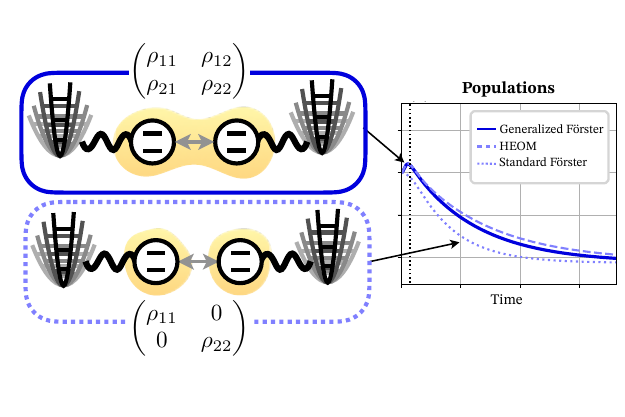}
\end{center}
\end{abstract}

\maketitle


In physical chemistry and some biological disciplines, Förster resonance energy transfer (FRET) is famously used as a ruler to estimate molecular distances \cite{CantorSchimmel}. FRET can play this role because of its specific distance dependence owing to the dipole-dipole character of the resonance intermolecular coupling. Leaving aside the nature of the coupling, FRET theory belongs to the same family of theories as Marcus electron transfer theory \cite{CantorSchimmel}, both of which follow straightforwardly from the Fermi golden rule \cite{MayKuehnBook}. In its original context of incoherent transfer, the rates are assumed to be slow with respect to the reorganization of the internal degrees of freedom (DOF) of the molecules and their environment. With the advent of ultrafast spectroscopy and its application to excitation energy transfer in pigment--protein complexes of photosynthesis, the Förster theory has become a robust first choice for the description of the excitation energy transfer between weakly coupled chromophores (see, e.g. \cite{zigmantas_two-dimensional_2006}). In this new context, the theory had to be extended to account for transfer between weakly coupled units of mutually strongly coupled chromophores, resulting in the so-called Multichromophoric Förster theory (see e.g. \cite{Sumi1999, Jang2004_MC-FRET}). Also, to accommodate the ultrafast time scale, i.e., to relax the condition of fast internal reorganization, the Nonequilibrium Förster theory was developed in ref. \cite{seibt_ultrafast_2017}, in which the bath reorganization dynamics is accounted for. 

Ultrafast laser pulse excitation of a multi-chromophoric system leads inevitably to a coherent initial state. However, no systematic treatment of decoherence is provided within the traditional Förster theory, and the rates of decoherence are typically postulated phenomenologically to satisfy a strict density matrix positivity condition embedded, e.g., within the Lindblad form \cite{lindblad1976,BreuerBook}. This requires the coherence element $\rho_{ab}=\mel{a}{\hat{\rho}}{b}$ to decay with one-half of the depopulation rates of each of the states $|a\rangle$ and $|b\rangle$ (the {\it one-half rule}). 
In this Letter, we introduce a general technique to derive an equation of motion (EM) for the  reduced density matrix (RDM) of an open quantum system, enabling us to derive a complete theory of Förster type that also rigorously includes decoherence dynamics. This makes the proposed theory truly suitable for the description of ultrafast optical experiments. The technique is based on an exact decomposition of the composite statistical operator into the system and bath components, together with a physically motivated time-dependent {\it ansatz} for the bath components. 
%
After laying out the theory for a general multistate molecular open quantum system, we study the dynamics of a donor--acceptor system and compare the results obtained from different types of master equations to the numerically exact solution of Hierarchical Equations of Motion (HEOM) \cite{Ishizaki2009c}. We identify the region of parameters in which different versions of the Förster theory can be considered valid. Most notably, we find a very high degree of agreement between the present generalized Förster theory and the exact dynamics during the initial transient time window of coherent dynamics. In contrast to the traditional weak intramolecular coupling theories, our formulation is well-defined also in the limit of the vanishing system--bath coupling for any value of the intramolecular coupling.
%

\emph{The system.---} Let us consider the Frenkel exciton Hamiltonian found in the theory of protein-embedded molecular aggregates (see, e.g. \cite{ValkunasMancalBook, Renger2013a}).
We divide it into four convenient components, $\hat{H} = \hat{H}_S + \hat{H}_B + \hat{V}_{S-S} + \hat{V}_{S-B}$.
The molecular system, $\hat{H}_S$, is made up of $N$ two-level molecules, $\hat{H}_a= \epsilon_{a}^{(g)}\ketbra{g_a}{g_a}+ \epsilon^{(e)}_a \ketbra{e_a}{e_a}$, where $\ket{g_a}$ and $\ket{e_a}$ represent the molecular electronic ground and excited states, respectively.
We consider the single-excitation approximation as appropriate for energy-transfer problems in photosynthesis, and define the collective single-excitation states $\ket{a} = \prod_{b\neq a}^{N}\ket{g_b} \ket{e_a}$ and ground state $|g\rangle = \prod_a |g_a\rangle$.
The system Hamiltonian then reads as $\hat{H}_S = \varepsilon_g \ketbra{g}{g} + \sum_{a=1}^N \varepsilon_a \ketbra{a}{a}$, with $\varepsilon_a = \epsilon^{(e)}_a$ and $\varepsilon_g=\sum_a \epsilon_g^{(a)}$ set to zero.
Electronic transitions on individual molecules interact through electrostatic resonance interaction $\hat{V}_{S-S} = \sum_{ab}J_{ab}\ketbra{a}{b}$. Their energy gaps are modulated by the pure dephasing interaction with the protein DOF and the intramolecular vibrational modes that together form their environment. 

When the molecules are in their electronic ground states, the environment (also referred to as the bath) is represented by an infinite set of harmonic modes with the Hamiltonian $\hat{H}_B = \frac{1}{2} \sum_{ak} \hbar\omega_{a,k}(\hat{p}_{a,k}^{2}+\hat{q}_{a,k}^{2})$. Upon electronic excitation, the pure dephasing interaction between electronic and bath DOF amounts to a potential energy surface shift $d_{b,k}$ such that the bath Hamiltonian experienced by the electronic excited state reads $\hat{H}_B^{(a)}=\frac{1}{2}\sum_{bk} \hbar\omega_{b,k}(\hat{p}_{b,k}^{2}+(\hat{q}_{b,k}-\delta_{ab}d_{b,k})^{2})$. 
The Hamiltonian $\hat{H}_0 = (\varepsilon_g + \hat{H}_B) \ketbra{g}{g} + \sum_a (\varepsilon_a + \hat{H}_B^{(a)})\ketbra{a}{a}$ can thus be described as independent levels coupled to different baths \cite{MahanBook, chenu2019a}, or equivalently, as $\hat{H}_0 = \hat{H}_S + \hat{H}_B + \hat{V}_{S-B}$, where the effective interaction $\hat{V}_{S-B} = \sum_a \hat{X}_a\ketbra{a}{a}$ is given by the so-called energy-gap operator $\ha X_a= -\sum_k \hbar \omega_{a,k} d_{a,k} \hat{q}_{a,k}$, linear in the bath coordinates.
The excited state energies of the aggregate molecules are each renormalized by the so-called reorganization energy $\lambda_a=\frac{1}{2}\sum_k \hbar\omega_{a,k} d_{a,k}^2$ and $\hat{H}_B^{(a)} = \hat{H}_B + \lambda_a + \hat{X}_a$.

The system--bath interaction is characterized by the so-called energy gap correlation functions (EGCFs), $C_{ab}(t)=\Tr_{B}\{\hat{X}_a(t)\hat{X}_b\hat{w}_{eq}\}$, where $\hat{w}_{eq}$ is the statistical operator for the bath at thermal equilibrium with respect to the molecular electronic ground state. The function $C_{ab}(t)$ measures the correlations of energy gap fluctuations. For photosynthetic systems, the fluctuations on different molecules are mostly uncorrelated, and  $C_{ab}(t)=C_{aa}(t)\delta_{ab}$ is a good approximation \cite{Olbrich2011b,Renger2012a}. The first and second integrals of $C_{ab}(t)$, i.e., $\dot{g}_{ab}(t)=\int_{0}^{t}{\rm d}\tau C_{ab}(\tau)$ and  the line-shape function $g_{ab}(t)=\int_{0}^{t}{\rm d}\tau\int_{0}^{\tau}{\rm d}\tau^{\prime} C_{ab}(\tau^{\prime})$, play an important role in the theory of spectroscopy and will be used to express our results in this paper. 

\emph{General second-order equation of motion.---} We will now present the equations governing the time evolution of the open quantum system's RDM $\hat \rho(t)$. We start with the full system-bath statistical operator $\hat{W}(t)$ that evolves according to the Liouville-von Neumann equation, $\frac{\partial}{\partial t}\hat{W}(t)=-\frac{i}{\hbar}[\hat{H},\hat{W}(t)]$. We divide the total Hamiltonian as $\hat{H} = \hat{H}_0+ \hat{V}$, where $\hat{V}$ is a suitably chosen weak perturbation. In the context of Förster resonance energy transfer \cite{forster_energiewanderung_1946, forster_zwischenmolekulare_1948}, the weak interaction is between the states of the system, i. e. $\hat{V}=\hat{V}_{S-S}$. The reference Hamiltonian $\hat{H}_0$ cannot be decomposed into a sum of $\hat{H}_S$ and $\hat{H}_B$, however, its form still allows us to evaluate the unperturbed dynamics exactly. 
Using the reference evolution operator $\hat{U}_0(t)=e^{- \frac{i}{\hbar} \hat{H}_0 t}$,
we derive the following identity [see Supplementary information (SI)], which forms the basis of our theoretical methodology:
\begin{eqnarray}
\label{eq:general_eom_w}
       \frac{\partial}{\partial t}\hat{W}(t)= - \frac{i}{\hbar}[\hat{V},\hat{U}_0(t)\hat{W}(0)\hat{U}_0^{\dagger}(t)]  -\frac{i}{\hbar}[\hat{H}_0,\hat{W}(t)] \nonumber  \\
   - \frac{ 1}{\hbar^2} 
    \! \int_{0}^{t}{\rm d}\tau\Big[\hat{V}, 
    \hat{U}_0(\tau)\big[\hat{V},
    \hat{W}(t{-}\tau)
    \big]\hat{U}^{\dagger}_0(\tau)\Big]. 
\end{eqnarray}
This equation of motion (EM) is exact and is nonlocal in time.
Tracing  over the bath DOF yields the dynamics relevant to the system of interest in terms of the RDM $\hat{\rho}(t)=\Tr_{B}\{\hat{W}(t)\}$.
However, since $\hat{W}(t)$ is surrounded by bath operators,
this last step is the main challenge --- and the central topic of this letter. 
It results in an approximate equation for the RDM that closely follows the general form: 
\begin{equation}
\label{eq:equation_foerster}
    \frac{\partial}{\partial t}\hat{\rho}(t) = \hat{I}(t) -i\mathcal{L}(t)\hat{\rho}(t)
    -\int_{0}^{t}{\rm d}\tau\; \mathcal{M}(t,\tau)\hat{\rho}(t-\tau).
\end{equation}
Here, $\hat{I}(t)$ is an operator depending on the initial condition $\hat{\rho}(0)$, $\mathcal{L}(t)$ is the Liouville superoperator capturing the unitary evolution and the decoherence due to pure dephasing interaction with the environment---with a time dependence that originates from the time-evolution of the bath state---and $\mathcal{M}(t, \tau)$ is the memory kernel superoperator, which captures the influence of the past on the present RDM. 

\emph{The bath state ansatz.---}
Our derivation of a general equation of motion relies on the observation that the total statistical operator can be factorized as 
\begin{equation}
\label{eq:Wt_fac}
    \hat{W}(t) = \sum_{ab}\rho_{ab}(t)\hat{w}_{ab}(t)\ketbra{a}{b},
\end{equation}
where the states $\{\ket{a}\}$ form a relevant physical basis for the system of interest (taken to be the site basis in Förster theory), $\rho_{ab}(t) = \bra{a}\hat{\rho}(t) \ket{b}=\Tr_{B}\{\hat{W}_{ab}(t)\}$, and where we have defined the relative bath state operators $\hat{w}_{ab}(t) \equiv \hat{W}_{ab}(t) / \rho_{ab}(t)$. 
To be able to employ the factorization, Eq. (\ref{eq:Wt_fac}), we need to approximate the relative bath operator $\hat{w}_{ab}(t)$ in a suitable way. The reference dynamics of perturbation theory is the evolution dictated by the Hamiltonian $\hat H_0$ alone. 
Taking such a zero's order (in $V_{S-S}$) evolution of the bath as the reference corresponds to the following time-dependent approximation for its  state operators, relative to the electronic states: 
\begin{equation}
\label{eq:ansatz}
\hat{w}_{ab}(t) = \frac{\hat{u}_a(t)\hat{w}_{\rm eq}\hat{u}^{\dagger}_{b}(t)}{\Tr_B\{\hat{u}_a(t)\hat{w}_{\eq}\hat{u}^{\dagger}_{b}(t)\}}.
\end{equation}
We will take this normalized density operator as an ansatz for the bath.
Here, the evolution operators $\hat{u}_a(t)\equiv e^{-i\hat{H}_B^{(a)}t/\hbar}$ describe the evolution of the bath on the potential energy surface (PES) of the electronic state $|a\rangle$ and $\hat{w}_\eq=e^{- \beta \hat{H}_B}/Z$ is the density operator of the bath in equilibrium with respect to the electronic ground state, $Z = \Tr_B(e^{- \beta \hat{H}_B})$ being the bath partition function. 
Although Eq. (\ref{eq:ansatz}) is a zeroth-order approximation, it conveniently corresponds to the bath state operators implicit to the non-equilibrium Förster theory \cite{seibt_ultrafast_2017}. 
%
%


 \emph{Equations of motion in Förster limit.---} We now use the approximation, Eq.  (\ref{eq:ansatz}), to derive the equations of motion for all elements of the RDM for a system of weakly coupled molecules. 
In the following, we calculate all the expected components of the master equation, Eq. (\ref{eq:equation_foerster}), using the cumulant expansion. The details of the derivations can be found in the SI.
 

\textit{(i) Initial term:}
The first term  on the r.h.s. of Eqs. (\ref{eq:general_eom_w})  and (\ref{eq:equation_foerster}) depends only on the initial state of the total system. We assume that the system is initially excited by a spectrally broad laser pulse. In the Condon approximation, this corresponds to $\hat{W}(0)=\hat{\rho}(0) \hat{w}_\eq$, with $\hat{\rho}(0)$  a possibly coherent superposition of electronic excited states. Such an initial condition is required for calculating non-linear optical response functions. The initial term can be expressed compactly by defining the dephasing operator 
$\hat{D}(t) \equiv \sum_{ab}D_{ab}(t) \ketbra{a}{b}$, 
with the elements
\begin{eqnarray}
\label{eq:G_nm}
   D_{ab}(t)&=&\Tr_{B}\{\hat{u}_{b}^{\dagger}(t)\hat{u}_{a}(t)\hat{w}_{eq}\} e^{-i\omega_{ab}t}\nonumber \\ 
   &=&e^{-g_{aa}(t)-g^{*}_{bb}(t)+2{\rm Re}(g_{ab}(t))-i\omega_{ab}t},
\end{eqnarray}
and the so-called Hadamard (element-wise) product denoted by $\circ$. 
We obtain 
\begin{equation}\label{eq:I_complete}
\ha I(t) = - \frac{i}{\hbar} [\hat V_{S-S}, \hat{D}(t)\circ \hat\rho(0)].
\end{equation}
This `initial term' connects the initial site populations with the coherence elements of the reduced statistical operator, and {\it vice versa}.
As expected, it redistributes the population during the time window in which it is non-zero, while keeping their total unchanged since $\sum_{a}I_{aa}(t)=0$. 

\textit{(ii) Hamiltonian term:}
The second term on the r.h.s. of Eqs. (\ref{eq:general_eom_w}) and (\ref{eq:equation_foerster}), is defined as the commutator of the total statistical operator $\hat{W}(t)$ with the Hamiltonian $\hat{H}_0$. This can be evaluated using our ansatz for the bath, Eq. (\ref{eq:ansatz}), as detailed in the SI, and 
written as
\begin{equation}
    {\cal L}(t)\hat{\rho}(t)=\hat{\Omega}(t)\circ \hat{\rho}(t),
\end{equation}
where the elements of the $\hat{\Omega}(t)$ 
operator are defined as 
\begin{eqnarray}
    \Omega_{ab}(t) &=&  i \dot{D}_{ab}(t) / D_{ab}(t) \\
   &=& \omega_{ab} - i\dot{g}_{aa}(t) -i \dot{g}_{bb}^*(t) + 2i{\rm Re}\dot{g}_{ab}(t). \nonumber
\end{eqnarray}
This result reproduces the pure dephasing dynamics of two uncoupled molecules. We assume $g_{ab}(t) = 0$ for $a \neq b$ in all calculations.  There is no contribution of the Hamiltonian term to population dynamics. 

\emph{(iii) Memory kernel:} The last terms of Eqs. (\ref{eq:general_eom_w}) and (\ref{eq:equation_foerster}) are the convolutions of the statistical operator with a memory kernel $\mathcal{M}$. The kernel contains two nested commutators of the total statistical operator with the interaction operators $\ha V$, i.e. four terms in total. Each term represents two actions of the interaction operator, at times $t-\tau$ and $t$. 
The kernel contributions can be evaluated by applying the factorization, Eq. (\ref{eq:Wt_fac}), to the total statistical operator and tracing the whole expression over the bath DOF. The kernel will be expressed in terms of the bath evolution correlators
$
     F_{cd}^{ad} (t, \tau) \equiv \Tr_{B}\{\hat{ u}_a(\tau) \ha w_{cd}(t-\tau) \hat{u}_d\dg(\tau)\}e^{-i\omega_{ad}\tau} , 
$
representing the history of bath time-evolution on different PES starting from the RDM element $\rho_{cd}$. 
Taking the approximate form, Eq. (\ref{eq:ansatz}), for the bath allows us to evaluate
the bath correlator as $ F_{cd}^{ad} (t, \tau)\approx f_{cd}^{ad}(t,\tau)e^{-i\omega_{ad}\tau}$ with:
\begin{equation}
\label{eq:correlator}
        f_{cd}^{ad}(t,\tau) = \frac{\Tr_{B}\{ \hat{u}_a(\tau)\hat{u}_c(t-\tau)\hat{w}_{eq}\hat{u}^{\dagger}_d(t) \}}{\Tr_{B}\{ \hat{u}_c(t-\tau)\hat{w}_{eq} \hat{u}^{\dagger}_d(t-\tau)\}}.
\end{equation}
The bath correlators fully characterize all integration kernels considered here and can be evaluated using the cumulant expansion technique. In order to avoid errors in this straightforward but tedious calculation, we evaluate Eq. \eqref{eq:correlator} using a symbolic module of our software package \textit{Quantarhei} \cite{Mancal2025_Quantarhei} which is based on Sympy \cite{Meurer2017SymPy}. This yields (see SI for details)
\begin{eqnarray}
\label{eq:fcda_cummulant}
f_{cd}^{ad}(t,\tau) &=& e^{ -g_{aa}(\tau)-g^{*}_{dd}(t)+g^{*}_{dd}(t-\tau)+g_{da}(\tau)+g^{*}_{da}(t)}  \nonumber \\
&\times& e^{ -g^{*}_{da}(t-\tau) -g_{ac}(t)+g_{ac}(\tau)+g_{ac}(t-\tau)} \nonumber \\
&\times& e^{-g_{dc}(\tau) + g_{dc}(t)-g_{dc}(t-\tau)}.
\end{eqnarray}  
The approximate correlator $f_{cd}^{ad}(t,\tau)$ conserves an important behavior of $F_{cd}^{ad}(t,\tau)$, namely that 
    $F_{ca}^{aa}(t,\tau)=f_{ca}^{aa}(t,\tau)=1$,
for all values of $c$ and $a$. An important special case is the expression where the diagram starts with $c=d$, i.e.
$f_{cc}^{ac}(t,\tau) = e^{ -g_{aa}(\tau)-g_{cc}(\tau)-ih_{cc}(t,\tau)}$, which appears in the expression for the population rates. Here, we define $h_{cc}(t,\tau)=2{\rm Im}[g_{cc}(t-\tau)-g_{cc}(t)]$.
 With the expression, Eq. \eqref{eq:fcda_cummulant}, for the correlators, Eq. \eqref{eq:correlator}, we can construct all the matrix elements entering the equation of motion for the reduced statistical operator. 

\emph{Non-equilibrium F\"orster rates in a donor-acceptor system ---}
For a simple molecular donor-acceptor system ($N=2$), all terms of Eq. (\ref{eq:general_eom_w}) can be easily written out and traced over the bath using the results and definitions listed above. We obtain EM for populations $\rho_{11}(t)$ and $\rho_{22}(t)$ that are independent of the coherence elements $\rho_{12}(t)$ and {\it vise versa}. The dimer problem is therefore inherently secular, with the exception of the dependence on the initial condition. The population part of the EM has the form
\begin{eqnarray}
\label{eq:F_time_non_loc}
\!\!\!\!&&\frac{\partial}{\partial t}\rho_{11}(t) = -\frac{i}{\hbar}J_{12}[D_{21}(t)\rho_{21}(0)-\rho_{12}(0)D_{12}(t)]  \\
    &-& \!\!\frac{|J_{12}|^2}{\hbar^2}2{\rm Re}\!\int\limits_{0}^{t}\!\!{\rm d}\tau
    e^{-g_{22}(\tau)-g_{11}(\tau)-ih_{11}(t,\tau)-i\omega_{21}\tau}\rho_{11}(t{-}\tau) \nonumber \\
    &+&\!\! \frac{|J_{12}|^2}{\hbar^2}2{\rm Re}\! \int\limits_{0}^{t}{\rm d}\tau
    e^{-g_{11}(\tau)-g_{22}(\tau)-ih_{22}(t,\tau)-i\omega_{12}\tau}\rho_{22}(t{-}\tau). \nonumber
\end{eqnarray}
Eq.~\eqref{eq:F_time_non_loc} is time-non-local and can be approximately time-localizated by approximating $\rho_{nn}(t-\tau)\approx\rho_{nn}(t)$. The resulting relaxation rates correspond exactly to the non-equilibrium Förster rates in Ref. \cite{seibt_ultrafast_2017}. Moreover, as the non-equilibrium Förster rates asymptotically converge to the standard Förster rates, the present theory leads to the canonical equilibrium  between the populations of the states $|1\rangle$ and $|2\rangle$ at $t\rightarrow \infty$. The initial condition term is transient due to the presence of the dephasing term $D_{ab}(t)$ of Eq. (\ref{eq:G_nm}) and it has already been identified in Ref. \cite{seibt_ultrafast_2017}. Here, we identify its crucial role for the short time dynamics. 

\emph{Decoherence dynamics.---}
In addition to the rate theory, our formulation also provides an expression for the evolution of coherences. Due to the analytical properties of the correlators, we have $f_{ca}^{aa}=1$, and the integral term of the equation lacks any dependence on the bath DOF. This gives the equation for the dephasing of coherence in the form of
\begin{eqnarray}
\label{eq:ForsterIntDiff}
    &&\frac{\partial}{\partial t}\rho_{12}(t)  = -\frac{i}{\hbar}J_{12}[\rho_{22}(0)-\rho_{11}(0)] -i\omega_{12}\rho_{12}(t)   \\ 
    &&-[\dot{g}_{11}(t)+\dot{g}^{*}_{22}(t)]\rho_{12}(t) 
     - \frac{4i|J_{12}|^2}{\hbar^2}\int_{0}^{t}{\rm d}\tau \textrm{Im}( \rho_{12}(\tau)). \nonumber 
\end{eqnarray}
To our knowledge, this is an entirely new form of dephasing equation valid in the Förster regime of energy transfer. Although Eq. (\ref{eq:ForsterIntDiff}) as a whole is not exact, the derivation of its surprising integral term does not contain any approximations. Eq. (\ref{eq:ForsterIntDiff}) remains valid even when the interaction with the bath is absent. By itself, it does not introduce any new dephasing. This implies that in the Förster regime of energy transfer, no transfer rate associated dephasing is present, and {\it the one-half rule does not apply}. Despite its integral form, the numerical solution of Eq. (\ref{eq:ForsterIntDiff}) is not more demanding than the solution of the time-local equation because the integral needs to be only incrementally updated in each time step. 
%
%
The non-integral terms of Eq. (\ref{eq:ForsterIntDiff}) have constant factors at large values of $t$. If the coherence is to become stationary as the system reaches equilibrium, the imaginary part of the coherence elements has to become zero. A non-zero constant imaginary part of the coherence element would lead to an ever increasing integral term and would contradict stationarity at the equilibrium. 

\emph{Quality of the generalized Förster theory.---} To verify that our methodology, denoted  here as the generalized Förster theory (gFT), indeed improves on the standard Förster theory \cite{ValkunasMancalBook} and even its recent non-equilibrium version (neqFT) \cite{seibt_ultrafast_2017}, we compare the results of Eqs. (\ref{eq:F_time_non_loc}) and (\ref{eq:ForsterIntDiff}) with the exact solution of the corresponding dynamics of HEOM calculated using QuTiP \cite{lambert_qutipv5_2026}.
We model the environment with a Drude-Lorentz spectral density, characterized by the reorganization energy $\lambda$, which is a measure of the system-bath coupling strength, and the bath correlation time $t_c$.
Using the Padé expansion, we obtain the correlation function $C_{ab}(t)$, which, by integration, provides $\dot{g}_{ab}(t)$ and $g_{ab}(t)$ as mentioned above.
For all the following results, we set $T = 277\text{K}$.
As a comparative metric for the results, we choose 
the trace distance $T(\ha \rho, \ha \sigma) = \frac{1}{2}\Tr(\sqrt{(\ha \rho - \ha \sigma)\dg (\ha \rho - \ha \sigma)})$  to define the maximum error with respect to HEOM over a time-interval $\varepsilon = \max_{t \in [t_0, t_1]}T(\hat \rho^{\rm xFT}(t), \hat \rho^{\rm HEOM}(t))$. We divide it into its population part $\varepsilon_{pop}$ and its coherence part $\varepsilon_{coh}$ (see SI). The neqFT time-dependent relaxation rates quickly converge to standard time-independent Förster rates, and correspondingly, we chose neqFT as the most favorable representation of the standard Förster energy transfer theory. We assume that it lacks the initial condition term. The gFT which contains the initial condition term can be time-localized yielding the neqFT rates. For the purpose of this paper, therefore, the time-localized gFT differs from the neqFT only by the presence of the initial condition term, and we can thus individually study the influence of time non-locality and the initial term on the quality of the simulated dynamics. 

\begin{figure}
    \includegraphics[width=\linewidth]{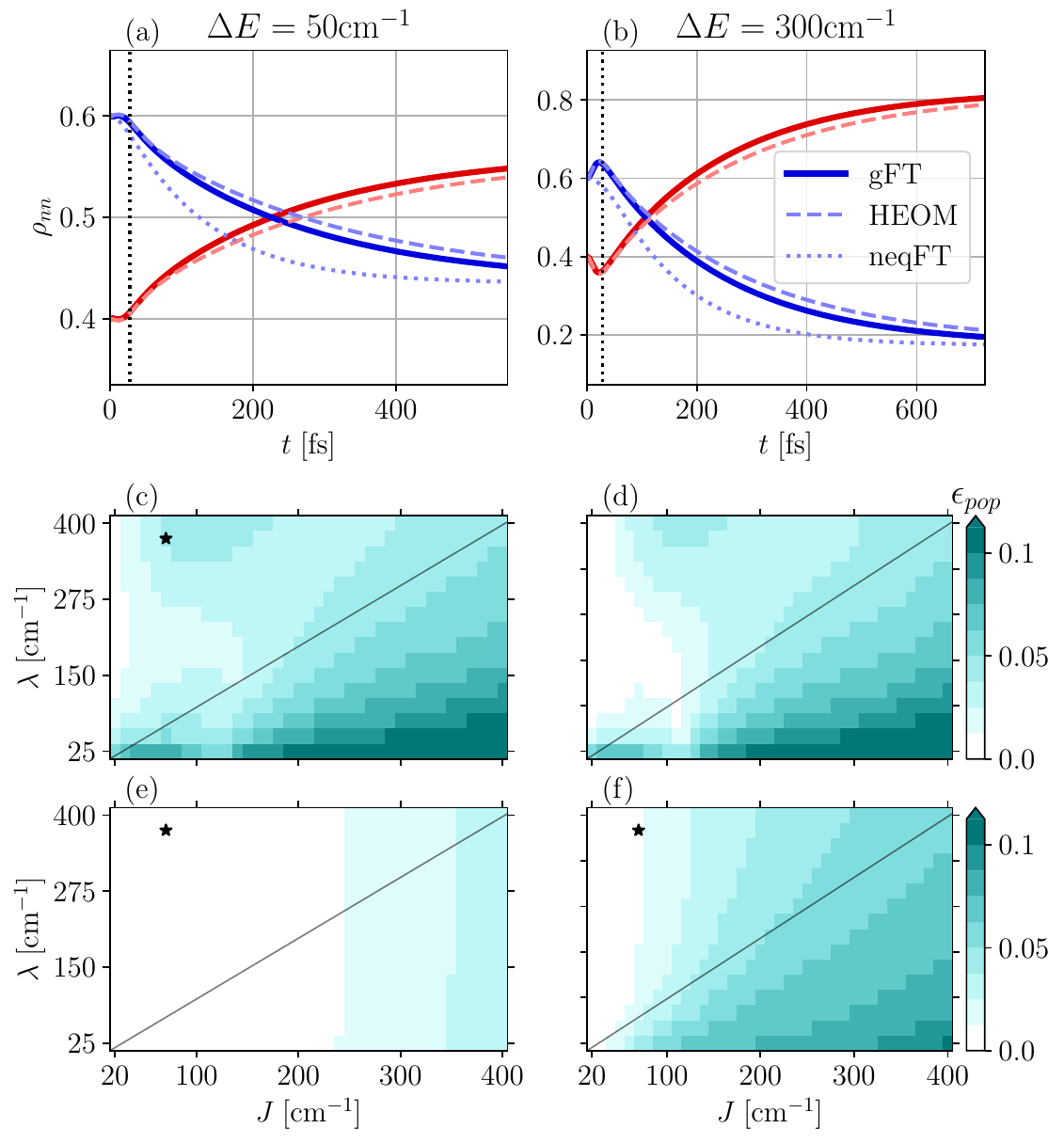}
    \caption{
        (a,b) Evolution of populations  in a donor-acceptor system as obtained from gFT, Eqs. \eqref{eq:F_time_non_loc} and \eqref{eq:ForsterIntDiff} (solid lines), from the exact numerical results using HEOM (dashed lines) and neqFT (dotted lines).
        The dashed black line indicates the bath correlation time $t_c = 30\; \mr{fs}$.
        (c-e) Short time ($t < t_c$) evaluation of maximum population-only trace distance $\varepsilon_{pop}$ for a range of $\lambda$ and $J$. The results from HEOM are compared with (c) neqFT, (d) time localized gFT and (e) full gFT.
        (f) The distance $\varepsilon_{pop}$ for gFT and HEOM calculated density matrices at intermediate times ($t_c < t < t_s$, where $t_s$ indicates the time at which equilibrium is reached).
        All diagrams (c-f) are for $\Delta E = 50 \mr{cm}^{-1}$
        with the star indicating the system parameters for the evolutions presented in (a,b): $J = 70\; \mr{cm}^{-1} \; \lambda=325\; \mr{cm}^{-1}$.
    The gray line indicates $J = \lambda$.
    }
    \label{fig:nr_01}
\end{figure}

In Fig. \ref{fig:nr_01}(a,b), we present typical population dynamics after a fully coherent initial condition $\rho_{11}(0)=0.4$, $\rho_{22}(0)=0.6$ and $\rho_{12}(0)=\sqrt{\rho_{11}(0)\rho_{22}(0)}$ calculated by HEOM, gFT and neqFT.
The parameters are listed in the figure caption.  We notice that neqFT lacks the initial fast population dynamics.
This is especially clear at a large energy gap of $\Delta E = 300\; \mr{cm^{-1}}$, exemplified in Fig. \ref{fig:nr_01}(b).
In Fig. \ref{fig:nr_01}(c-e) we present the maximum $\varepsilon_{pop}$ for gFT [including also its time-localized version (d)] and neqFT compared to HEOM, respectively, for a wide range of system parameters $\lambda$ and $J$ at short times ($t < t_c$).
The parameter region where gFT agrees with HEOM in the window of short times, i.e. when dynamics is influenced by initial coherence, far exceeds the expected region of weak resonance coupling.
The quality of agreement between gFT and HEOM depends, to some extent, only on the coupling $J$.
Comparing results in Fig. \ref{fig:nr_01}(d) and (e), shows that this is largely due to time non-local effects.
This is a clear improvement over neqFT, Fig. \ref{fig:nr_01}(c), which shows noticeable disagreement with the results of HEOM. For short time dynamics, combination of time non-local equations and the initial term yields significantly better results than the presence of the initial term alone. After the fast transient dynamics ($t>t_c)$, all versions of the dynamics are clearly rate dominated, and the gFT gives a rate that agrees very well with the effective rate corresponding to the HEOM dynamics. However, for a good agreement of the dynamics in this time window, it is crucial that the dynamics agree during the short time window.
As presented in Fig. \ref{fig:nr_01}(f), in the intermediate time window,  we find that the quality of the gFT dynamics depends non-linearly on both $\lambda$ and $J$ with increasing $\lambda$ improving the agreement for any value of $J$.
In the photosynthetic FMO complex that clearly shows signs of exciton delocalization, the maximum couplings are $J\sim 130$ cm$^{-1}$, depending on the parameterization \cite{Adolphs2006a}, while the typical value is around 20 or $30\; \mr{cm^{-1}}$.
The energy gaps are up to $400\; \mr{cm^{-1}}$.
Therefore, gFT would perform rather well in FMO for short and intermediate times with $\varepsilon_{pop} < 0.04$ for typical values of resonance coupling and reorganization energy. Note that, remarkably, the intermediate time error of gFT is comparable, and even smaller, than the short time error of the alternative Förster theories.

\begin{figure}
    \includegraphics[width=\linewidth]{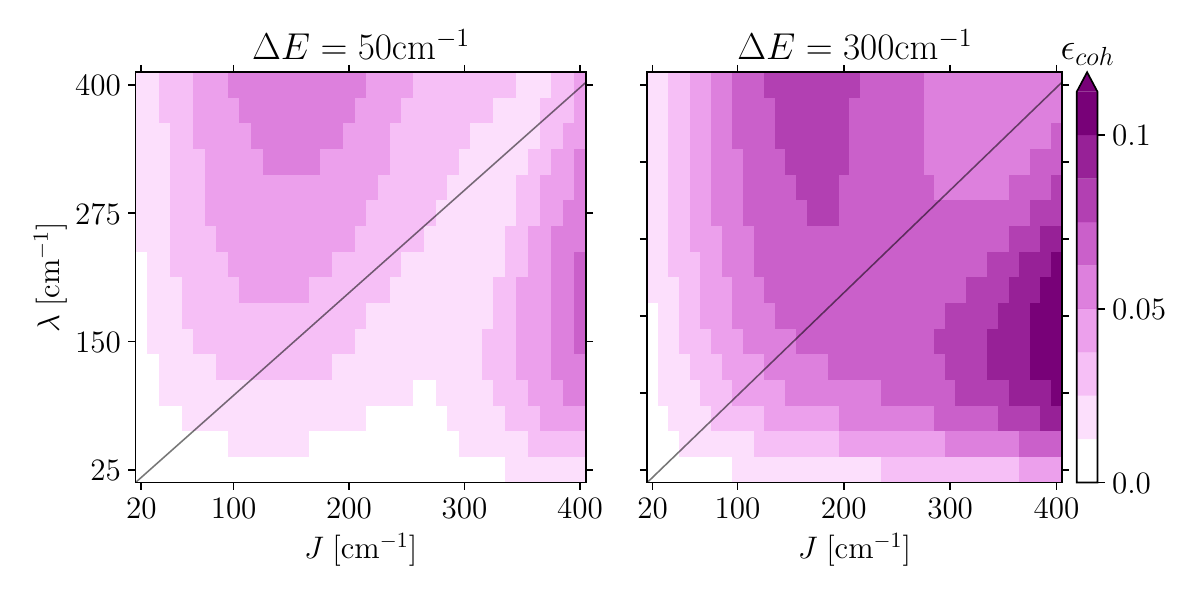}
    \caption{
        Short time ($t < t_c$) evaluation of coherence-only trace distance $\varepsilon_{coh}$ in a donor-acceptor system as obtained from the derived EM Eqs. \eqref{eq:F_time_non_loc}, \eqref{eq:ForsterIntDiff} and HEOM at two different energy gaps $\Delta E$.
        At the gray line $J = \lambda$.
    }
    \label{fig:nr_02}
\end{figure}

In Fig. \ref{fig:nr_02}, we present $\varepsilon_{coh}$ for the same range of parameters as discussed for populations.
It is clear that the  gFT describes the coherence evolution much less accurately than the populations, especially at longer times. This is to be expected for systems showing the effect of electronic state delocalization which translates into non-zero real values of long time coherence elements of the RDM when represented in the basis of local molecular states. This effect is not captured by the present theory at all. The detailed analysis of the coherence evolutions can be found in the SI, confirming the agreement of the gFT and HEOM for the coherence evolution at very short times, good agreement of the imaginary part of the coherence elements, and the failure of the gFT to capture of the non-zero real parts of the RDM at long times. 
Interestingly, the error depends non-trivially on the relative strength of the resonance and the system-bath couplings and is pronounced especially at larger energy gaps. The description of coherence by gFT is very good for both limits of small $\lambda$ and small $J$. 

\emph{Conclusions ---}
In this letter, we derived a complete second-order reduced density matrix theory valid in the parameter regime characterized by the weak intermolecular coupling and the strong system--bath coupling (the Förster theory regime).
We found that coherence in the initial condition rapidly redistributes populations and thus causes a lasting mismatch between exact dynamics and the standard Förster theory.
Our theory provides a novel form of decoherence equation that accurately captures the initial coherent dynamics of the system for the parameter range exceeding the weak resonance coupling regime.
Besides the weak coupling, the present theory is also exact for vanishing system-bath coupling and an arbitrary strength of resonance coupling.
We find that all variants of the Förster theories predict similar energy transfer rates that compare well to the dynamics obtained from HEOM in an extensive region of parameters. However, capturing the initial coherent dynamics correctly, by including the initial correlation term and time-non-local form of the relaxation, is crucial for agreement with the exact dynamics, even in propagation times when the dynamics is dominated by the rate process.
The technique we introduced in this letter for deriving master equations for the reduced density matrix can be combined with better approximations to the bath evolution to further improve the agreement with exact dynamics and to adapt Förster theory for different spectroscopic situations including multi-dimensional spectroscopies.

\begin{acknowledgments}
We thank Alessandro Manacorda and Veljko Jankovic for insightful discussions and Neill Lambert for helping with the HEOM simulations. This work was partially funded by the Luxembourg National Research Fund (FNR, Attract grant 15382998). TM acknowledges funding support from the Czech Science Foundation via grant no. 22-26376S.
\end{acknowledgments}


\bibliography{main, references, tomas_refs, max_refs}

\begin{thebibliography}{24}%
\makeatletter
\providecommand \@ifxundefined [1]{%
 \@ifx{#1\undefined}
}%
\providecommand \@ifnum [1]{%
 \ifnum #1\expandafter \@firstoftwo
 \else \expandafter \@secondoftwo
 \fi
}%
\providecommand \@ifx [1]{%
 \ifx #1\expandafter \@firstoftwo
 \else \expandafter \@secondoftwo
 \fi
}%
\providecommand \natexlab [1]{#1}%
\providecommand \enquote  [1]{``#1''}%
\providecommand \bibnamefont  [1]{#1}%
\providecommand \bibfnamefont [1]{#1}%
\providecommand \citenamefont [1]{#1}%
\providecommand \href@noop [0]{\@secondoftwo}%
\providecommand \href [0]{\begingroup \@sanitize@url \@href}%
\providecommand \@href[1]{\@@startlink{#1}\@@href}%
\providecommand \@@href[1]{\endgroup#1\@@endlink}%
\providecommand \@sanitize@url [0]{\catcode `\\12\catcode `\$12\catcode
  `\&12\catcode `\#12\catcode `\^12\catcode `\_12\catcode `\%12\relax}%
\providecommand \@@startlink[1]{}%
\providecommand \@@endlink[0]{}%
\providecommand \url  [0]{\begingroup\@sanitize@url \@url }%
\providecommand \@url [1]{\endgroup\@href {#1}{\urlprefix }}%
\providecommand \urlprefix  [0]{URL }%
\providecommand \Eprint [0]{\href }%
\providecommand \doibase [0]{https://doi.org/}%
\providecommand \selectlanguage [0]{\@gobble}%
\providecommand \bibinfo  [0]{\@secondoftwo}%
\providecommand \bibfield  [0]{\@secondoftwo}%
\providecommand \translation [1]{[#1]}%
\providecommand \BibitemOpen [0]{}%
\providecommand \bibitemStop [0]{}%
\providecommand \bibitemNoStop [0]{.\EOS\space}%
\providecommand \EOS [0]{\spacefactor3000\relax}%
\providecommand \BibitemShut  [1]{\csname bibitem#1\endcsname}%
\let\auto@bib@innerbib\@empty
\bibitem [{\citenamefont {Cantor}\ and\ \citenamefont
  {Schimmel}(2001)}]{CantorSchimmel}%
  \BibitemOpen
  \bibfield  {author} {\bibinfo {author} {\bibfnamefont {C.~R.}\ \bibnamefont
  {Cantor}}\ and\ \bibinfo {author} {\bibfnamefont {P.~R.}\ \bibnamefont
  {Schimmel}},\ }\href@noop {} {\emph {\bibinfo {title} {Biophysical Chemistry,
  Part II}}}\ (\bibinfo  {publisher} {{W. H. Freeman and co.}},\ \bibinfo
  {address} {{New York}},\ \bibinfo {year} {2001})\BibitemShut {NoStop}%
\bibitem [{\citenamefont {May}\ and\ \citenamefont
  {K{\"u}hn}(2001)}]{MayKuehnBook}%
  \BibitemOpen
  \bibfield  {author} {\bibinfo {author} {\bibfnamefont {V.}~\bibnamefont
  {May}}\ and\ \bibinfo {author} {\bibfnamefont {O.}~\bibnamefont {K{\"u}hn}},\
  }\href@noop {} {\emph {\bibinfo {title} {Charge and Energy Transfer Dynamics
  in Molecular Systems}}}\ (\bibinfo  {publisher} {{Wiley-VCH}},\ \bibinfo
  {address} {{Berlin}},\ \bibinfo {year} {2001})\BibitemShut {NoStop}%
\bibitem [{\citenamefont {Zigmantas}\ \emph {et~al.}(2006)\citenamefont
  {Zigmantas}, \citenamefont {Read}, \citenamefont {Man\v{c}al}, \citenamefont
  {Brixner}, \citenamefont {Gardiner}, \citenamefont {Cogdell},\ and\
  \citenamefont {Fleming}}]{zigmantas_two-dimensional_2006}%
  \BibitemOpen
  \bibfield  {author} {\bibinfo {author} {\bibfnamefont {D.}~\bibnamefont
  {Zigmantas}}, \bibinfo {author} {\bibfnamefont {E.~L.}\ \bibnamefont {Read}},
  \bibinfo {author} {\bibfnamefont {T.}~\bibnamefont {Man\v{c}al}}, \bibinfo
  {author} {\bibfnamefont {T.}~\bibnamefont {Brixner}}, \bibinfo {author}
  {\bibfnamefont {A.~T.}\ \bibnamefont {Gardiner}}, \bibinfo {author}
  {\bibfnamefont {R.~J.}\ \bibnamefont {Cogdell}},\ and\ \bibinfo {author}
  {\bibfnamefont {G.~R.}\ \bibnamefont {Fleming}},\ }\bibfield  {title}
  {\bibinfo {title} {Two-dimensional electronic spectroscopy of the
  {B800}--{B820} light-harvesting complex},\ }\href
  {https://doi.org/10.1073/pnas.0602961103} {\bibfield  {journal} {\bibinfo
  {journal} {Proceedings of the National Academy of Sciences}\ }\textbf
  {\bibinfo {volume} {103}},\ \bibinfo {pages} {12672} (\bibinfo {year}
  {2006})}\BibitemShut {NoStop}%
\bibitem [{\citenamefont {Sumi}(1999)}]{Sumi1999}%
  \BibitemOpen
  \bibfield  {author} {\bibinfo {author} {\bibfnamefont {H.}~\bibnamefont
  {Sumi}},\ }\bibfield  {title} {\bibinfo {title} {Theory on rates of
  excitation-energy transfer between molecular aggregates through distributed
  transition dipoles with application to the antenna system in bacterial
  photosynthesis},\ }\href {https://doi.org/10.1021/jp983477u} {\bibfield
  {journal} {\bibinfo  {journal} {J. Phys. Chem. B}\ }\textbf {\bibinfo
  {volume} {103}},\ \bibinfo {pages} {252--260} (\bibinfo {year}
  {1999})}\BibitemShut {NoStop}%
\bibitem [{\citenamefont {Jang}\ \emph {et~al.}(2004)\citenamefont {Jang},
  \citenamefont {Newton},\ and\ \citenamefont {Silbey}}]{Jang2004_MC-FRET}%
  \BibitemOpen
  \bibfield  {author} {\bibinfo {author} {\bibfnamefont {S.}~\bibnamefont
  {Jang}}, \bibinfo {author} {\bibfnamefont {M.~D.}\ \bibnamefont {Newton}},\
  and\ \bibinfo {author} {\bibfnamefont {R.~J.}\ \bibnamefont {Silbey}},\
  }\bibfield  {title} {\bibinfo {title} {Multichromophoric f\"orster resonance
  energy transfer},\ }\href {https://doi.org/10.1103/PhysRevLett.92.218301}
  {\bibfield  {journal} {\bibinfo  {journal} {Phys. Rev. Lett.}\ }\textbf
  {\bibinfo {volume} {92}},\ \bibinfo {pages} {218301} (\bibinfo {year}
  {2004})}\BibitemShut {NoStop}%
\bibitem [{\citenamefont {Seibt}\ and\ \citenamefont
  {Man\v{c}al}(2017)}]{seibt_ultrafast_2017}%
  \BibitemOpen
  \bibfield  {author} {\bibinfo {author} {\bibfnamefont {J.}~\bibnamefont
  {Seibt}}\ and\ \bibinfo {author} {\bibfnamefont {T.}~\bibnamefont
  {Man\v{c}al}},\ }\bibfield  {title} {\bibinfo {title} {Ultrafast energy transfer
  with competing channels: {Non}-equilibrium {F\"orster} and {Modified}
  {Redfield} theories},\ }\href {https://doi.org/10.1063/1.4981523} {\bibfield
  {journal} {\bibinfo  {journal} {The Journal of Chemical Physics}\ }\textbf
  {\bibinfo {volume} {146}},\ \bibinfo {pages} {174109} (\bibinfo {year}
  {2017})}\BibitemShut {NoStop}%
\bibitem [{\citenamefont {Lindblad}(1976)}]{lindblad1976}%
  \BibitemOpen
  \bibfield  {author} {\bibinfo {author} {\bibfnamefont {G.}~\bibnamefont
  {Lindblad}},\ }\bibfield  {title} {\bibinfo {title} {On the generators of
  quantum dynamical semigroups},\ }\href {https://doi.org/10.1007/BF01608499}
  {\bibfield  {journal} {\bibinfo  {journal} {Comm. Math. Phys.}\ }\textbf
  {\bibinfo {volume} {48}},\ \bibinfo {pages} {119} (\bibinfo {year}
  {1976})}\BibitemShut {NoStop}%
\bibitem [{\citenamefont {Breuer}\ and\ \citenamefont
  {Petruccione}(2007)}]{BreuerBook}%
  \BibitemOpen
  \bibfield  {author} {\bibinfo {author} {\bibfnamefont {H.-P.}\ \bibnamefont
  {Breuer}}\ and\ \bibinfo {author} {\bibfnamefont {F.}~\bibnamefont
  {Petruccione}},\ }\href@noop {} {\emph {\bibinfo {title} {The Theory of Open
  Quantum Systems}}}\ (\bibinfo  {publisher} {{Oxford University Press}},\
  \bibinfo {year} {2007})\BibitemShut {NoStop}%
\bibitem [{\citenamefont {Ishizaki}\ and\ \citenamefont
  {Fleming}(2009)}]{Ishizaki2009c}%
  \BibitemOpen
  \bibfield  {author} {\bibinfo {author} {\bibfnamefont {A.}~\bibnamefont
  {Ishizaki}}\ and\ \bibinfo {author} {\bibfnamefont {G.~R.}\ \bibnamefont
  {Fleming}},\ }\bibfield  {title} {\bibinfo {title} {Unified treatment of
  quantum coherent and incoherent hopping dynamics in electronic energy
  transfer: {{Reduced}} hierarchy equation approach},\ }\href
  {https://doi.org/10.1063/1.3155372} {\bibfield  {journal} {\bibinfo
  {journal} {J. Chem. Phys.}\ }\textbf {\bibinfo {volume} {130}},\ \bibinfo
  {pages} {234111} (\bibinfo {year} {2009})}\BibitemShut {NoStop}%
\bibitem [{\citenamefont {Valkunas}\ \emph {et~al.}(2013)\citenamefont
  {Valkunas}, \citenamefont {Abramavi{\v c}ius},\ and\ \citenamefont {Man{\v
  c}al}}]{ValkunasMancalBook}%
  \BibitemOpen
  \bibfield  {author} {\bibinfo {author} {\bibfnamefont {L.}~\bibnamefont
  {Valkunas}}, \bibinfo {author} {\bibfnamefont {D.}~\bibnamefont {Abramavi{\v
  c}ius}},\ and\ \bibinfo {author} {\bibfnamefont {T.}~\bibnamefont {Man{\v
  c}al}},\ }\href@noop {} {\emph {\bibinfo {title} {Molecular Excitation
  Dynamics and Relaxation: Quantum Theory and Spectroscopy}}}\ (\bibinfo
  {publisher} {{Wiley-VCH}},\ \bibinfo {address} {{Weinheim}},\ \bibinfo {year}
  {2013})\BibitemShut {NoStop}%
\bibitem [{\citenamefont {Renger}\ and\ \citenamefont
  {M\"uh}(2013)}]{Renger2013a}%
  \BibitemOpen
  \bibfield  {author} {\bibinfo {author} {\bibfnamefont {T.}~\bibnamefont
  {Renger}}\ and\ \bibinfo {author} {\bibfnamefont {F.}~\bibnamefont {M\"uh}},\
  }\href {https://doi.org/10.1039/c3cp43439g} {\bibinfo {title} {Understanding
  photosynthetic light-harvesting: A bottom up theoretical approach}} (\bibinfo
  {year} {2013})\BibitemShut {NoStop}%
\bibitem [{\citenamefont {Mahan}(2000)}]{MahanBook}%
  \BibitemOpen
  \bibfield  {author} {\bibinfo {author} {\bibfnamefont {G.~D.}\ \bibnamefont
  {Mahan}},\ }\href@noop {} {\emph {\bibinfo {title} {Many-{{Particle
  Physics}}}}},\ \bibinfo {edition} {3rd}\ ed.\ (\bibinfo  {publisher} {{Kluwer
  Academic Publishers}},\ \bibinfo {year} {2000})\BibitemShut {NoStop}%
\bibitem [{\citenamefont {Chenu}\ \emph {et~al.}(2019)\citenamefont {Chenu},
  \citenamefont {Shiau},\ and\ \citenamefont {Combescot}}]{chenu2019a}%
  \BibitemOpen
  \bibfield  {author} {\bibinfo {author} {\bibfnamefont {A.}~\bibnamefont
  {Chenu}}, \bibinfo {author} {\bibfnamefont {S.-Y.}\ \bibnamefont {Shiau}},\
  and\ \bibinfo {author} {\bibfnamefont {M.}~\bibnamefont {Combescot}},\
  }\bibfield  {title} {\bibinfo {title} {Two-level system coupled to phonons:
  {{Full}} analytical solution},\ }\href
  {https://doi.org/10.1103/PhysRevB.99.014302} {\bibfield  {journal} {\bibinfo
  {journal} {Phys. Rev. B}\ }\textbf {\bibinfo {volume} {99}},\ \bibinfo
  {pages} {014302} (\bibinfo {year} {2019})}\BibitemShut {NoStop}%
\bibitem [{\citenamefont {Olbrich}\ \emph {et~al.}(2011)\citenamefont
  {Olbrich}, \citenamefont {Str{\"u}mpfer}, \citenamefont {Schulten},\ and\
  \citenamefont {Kleinekath{\"o}fer}}]{Olbrich2011b}%
  \BibitemOpen
  \bibfield  {author} {\bibinfo {author} {\bibfnamefont {C.}~\bibnamefont
  {Olbrich}}, \bibinfo {author} {\bibfnamefont {J.}~\bibnamefont
  {Str{\"u}mpfer}}, \bibinfo {author} {\bibfnamefont {K.}~\bibnamefont
  {Schulten}},\ and\ \bibinfo {author} {\bibfnamefont {U.}~\bibnamefont
  {Kleinekath{\"o}fer}},\ }\bibfield  {title} {\bibinfo {title} {Quest for
  spatially correlated fluctuations in the {{FMO}} light-harvesting complex},\
  }\href@noop {} {\bibfield  {journal} {\bibinfo  {journal} {J. Phys. Chem. B}\
  }\textbf {\bibinfo {volume} {115}},\ \bibinfo {pages} {758} (\bibinfo {year}
  {2011})}\BibitemShut {NoStop}%
\bibitem [{\citenamefont {Renger}\ \emph {et~al.}(2012)\citenamefont {Renger},
  \citenamefont {Klinger}, \citenamefont {Steinecker}, \citenamefont {Busch},
  \citenamefont {Numata},\ and\ \citenamefont {M\"uh}}]{Renger2012a}%
  \BibitemOpen
  \bibfield  {author} {\bibinfo {author} {\bibfnamefont {T.}~\bibnamefont
  {Renger}}, \bibinfo {author} {\bibfnamefont {A.}~\bibnamefont {Klinger}},
  \bibinfo {author} {\bibfnamefont {F.}~\bibnamefont {Steinecker}}, \bibinfo
  {author} {\bibfnamefont {M.~S.~A.}\ \bibnamefont {Busch}}, \bibinfo {author}
  {\bibfnamefont {J.}~\bibnamefont {Numata}},\ and\ \bibinfo {author}
  {\bibfnamefont {F.}~\bibnamefont {M\"uh}},\ }\bibfield  {title} {\bibinfo
  {title} {Normal mode analysis of the spectral density of the
  fenna-matthews-olson light-harvesting protein: How the protein dissipates the
  excess energy of excitons},\ }\href {https://doi.org/10.1021/jp3094935}
  {\bibfield  {journal} {\bibinfo  {journal} {Journal of Physical Chemistry B}\
  }\textbf {\bibinfo {volume} {116}},\ \bibinfo {pages} {14565} (\bibinfo
  {year} {2012})}\BibitemShut {NoStop}%
\bibitem [{\citenamefont {Forster}(1946)}]{forster_energiewanderung_1946}%
  \BibitemOpen
  \bibfield  {author} {\bibinfo {author} {\bibfnamefont {T.}~\bibnamefont
  {Forster}},\ }\bibfield  {title} {\bibinfo {title} {Energiewanderung und
  {Fluoreszenz}},\ }\href {https://doi.org/10.1007/BF00585226} {\bibfield
  {journal} {\bibinfo  {journal} {Die Naturwissenschaften}\ }\textbf {\bibinfo
  {volume} {33}},\ \bibinfo {pages} {166} (\bibinfo {year} {1946})}\BibitemShut
  {NoStop}%
\bibitem [{\citenamefont {F\"orster}(1948)}]{forster_zwischenmolekulare_1948}%
  \BibitemOpen
  \bibfield  {author} {\bibinfo {author} {\bibfnamefont {T.}~\bibnamefont
  {F\"orster}},\ }\bibfield  {title} {\bibinfo {title} {Zwischenmolekulare
  {Energiewanderung} und {Fluoreszenz}},\ }\href
  {https://doi.org/10.1002/andp.19484370105} {\bibfield  {journal} {\bibinfo
  {journal} {Annalen der Physik}\ }\textbf {\bibinfo {volume} {437}},\ \bibinfo
  {pages} {55} (\bibinfo {year} {1948})}\BibitemShut {NoStop}%
\bibitem [{\citenamefont {Man\v{c}al}\ and\ \citenamefont
  {contributors}(2025)}]{Mancal2025_Quantarhei}%
  \BibitemOpen
  \bibfield  {author} {\bibinfo {author} {\bibfnamefont {T.}~\bibnamefont
  {Man\v{c}al}}\ and\ \bibinfo {author} {\bibnamefont {contributors}},\ }\href
  {https://github.com/tmancal74/quantarhei} {\emph {\bibinfo {title}
  {{Quantarhei:} Open quantum-system simulator for molecular aggregates}}},\
  \bibinfo {organization} {Quantarhei Development Team} (\bibinfo {year}
  {2025}),\ \bibinfo {note} {commit/tag or release version used for
  simulations}\BibitemShut {NoStop}%
\bibitem [{\citenamefont {Meurer}\ \emph {et~al.}(2017)\citenamefont {Meurer},
  \citenamefont {Smith}, \citenamefont {Paprocki}, \citenamefont {\v{C}ert\'ik},
  \citenamefont {Kirpichev}, \citenamefont {Rocklin}, \citenamefont {Kumar},
  \citenamefont {Ivanov}, \citenamefont {Moore}, \citenamefont {Singh},
  \citenamefont {Rathnayake}, \citenamefont {Vig}, \citenamefont {Granger},
  \citenamefont {M\"uller}, \citenamefont {Bonazzi}, \citenamefont {Gupta},
  \citenamefont {Vats}, \citenamefont {Johansson}, \citenamefont {Pedregosa},
  \citenamefont {Curry}, \citenamefont {Terrel}, \citenamefont {Rou\v{c}ka},
  \citenamefont {Saboo}, \citenamefont {Fernando}, \citenamefont {Kulal},
  \citenamefont {Cimrman},\ and\ \citenamefont {Scopatz}}]{Meurer2017SymPy}%
  \BibitemOpen
  \bibfield  {author} {\bibinfo {author} {\bibfnamefont {A.}~\bibnamefont
  {Meurer}}, \bibinfo {author} {\bibfnamefont {C.~P.}\ \bibnamefont {Smith}},
  \bibinfo {author} {\bibfnamefont {M.}~\bibnamefont {Paprocki}}, \bibinfo
  {author} {\bibfnamefont {O.}~\bibnamefont {\v{C}ert\'ik}}, \bibinfo {author}
  {\bibfnamefont {S.~B.}\ \bibnamefont {Kirpichev}}, \bibinfo {author}
  {\bibfnamefont {M.}~\bibnamefont {Rocklin}}, \bibinfo {author} {\bibfnamefont
  {A.}~\bibnamefont {Kumar}}, \bibinfo {author} {\bibfnamefont
  {S.}~\bibnamefont {Ivanov}}, \bibinfo {author} {\bibfnamefont {J.~K.}\
  \bibnamefont {Moore}}, \bibinfo {author} {\bibfnamefont {S.}~\bibnamefont
  {Singh}}, \bibinfo {author} {\bibfnamefont {T.}~\bibnamefont {Rathnayake}},
  \bibinfo {author} {\bibfnamefont {S.}~\bibnamefont {Vig}}, \bibinfo {author}
  {\bibfnamefont {B.~E.}\ \bibnamefont {Granger}}, \bibinfo {author}
  {\bibfnamefont {R.~P.}\ \bibnamefont {M\"uller}}, \bibinfo {author}
  {\bibfnamefont {F.}~\bibnamefont {Bonazzi}}, \bibinfo {author} {\bibfnamefont
  {H.}~\bibnamefont {Gupta}}, \bibinfo {author} {\bibfnamefont
  {S.}~\bibnamefont {Vats}}, \bibinfo {author} {\bibfnamefont {F.}~\bibnamefont
  {Johansson}}, \bibinfo {author} {\bibfnamefont {F.}~\bibnamefont
  {Pedregosa}}, \bibinfo {author} {\bibfnamefont {M.~J.}\ \bibnamefont
  {Curry}}, \bibinfo {author} {\bibfnamefont {A.~R.}\ \bibnamefont {Terrel}},
  \bibinfo {author} {\bibfnamefont {V.}~\bibnamefont {Rou\v{c}ka}}, \bibinfo
  {author} {\bibfnamefont {A.}~\bibnamefont {Saboo}}, \bibinfo {author}
  {\bibfnamefont {I.}~\bibnamefont {Fernando}}, \bibinfo {author}
  {\bibfnamefont {S.}~\bibnamefont {Kulal}}, \bibinfo {author} {\bibfnamefont
  {R.}~\bibnamefont {Cimrman}},\ and\ \bibinfo {author} {\bibfnamefont
  {A.}~\bibnamefont {Scopatz}},\ }\bibfield  {title} {\bibinfo {title} {Sympy:
  symbolic computing in python},\ }\href {https://doi.org/10.7717/peerj-cs.103}
  {\bibfield  {journal} {\bibinfo  {journal} {PeerJ Computer Science}\ }\textbf
  {\bibinfo {volume} {3}},\ \bibinfo {pages} {e103} (\bibinfo {year}
  {2017})}\BibitemShut {NoStop}%
\bibitem [{\citenamefont {Lambert}\ \emph {et~al.}(2026)\citenamefont
  {Lambert}, \citenamefont {Gigu\`ere}, \citenamefont {Menczel}, \citenamefont
  {Li}, \citenamefont {Hopf}, \citenamefont {Su\'arez}, \citenamefont {Gali},
  \citenamefont {Lishman}, \citenamefont {Gadhvi}, \citenamefont {Agarwal},
  \citenamefont {Galicia}, \citenamefont {Shammah}, \citenamefont {Nation},
  \citenamefont {Johansson}, \citenamefont {Ahmed}, \citenamefont {Cross},
  \citenamefont {Pitchford},\ and\ \citenamefont
  {Nori}}]{lambert_qutipv5_2026}%
  \BibitemOpen
  \bibfield  {author} {\bibinfo {author} {\bibfnamefont {N.}~\bibnamefont
  {Lambert}}, \bibinfo {author} {\bibfnamefont {E.}~\bibnamefont {Gigu\`ere}},
  \bibinfo {author} {\bibfnamefont {P.}~\bibnamefont {Menczel}}, \bibinfo
  {author} {\bibfnamefont {B.}~\bibnamefont {Li}}, \bibinfo {author}
  {\bibfnamefont {P.}~\bibnamefont {Hopf}}, \bibinfo {author} {\bibfnamefont
  {G.}~\bibnamefont {Su\'arez}}, \bibinfo {author} {\bibfnamefont
  {M.}~\bibnamefont {Gali}}, \bibinfo {author} {\bibfnamefont {J.}~\bibnamefont
  {Lishman}}, \bibinfo {author} {\bibfnamefont {R.}~\bibnamefont {Gadhvi}},
  \bibinfo {author} {\bibfnamefont {R.}~\bibnamefont {Agarwal}}, \bibinfo
  {author} {\bibfnamefont {A.}~\bibnamefont {Galicia}}, \bibinfo {author}
  {\bibfnamefont {N.}~\bibnamefont {Shammah}}, \bibinfo {author} {\bibfnamefont
  {P.}~\bibnamefont {Nation}}, \bibinfo {author} {\bibfnamefont
  {J.}~\bibnamefont {Johansson}}, \bibinfo {author} {\bibfnamefont
  {S.}~\bibnamefont {Ahmed}}, \bibinfo {author} {\bibfnamefont
  {S.}~\bibnamefont {Cross}}, \bibinfo {author} {\bibfnamefont
  {A.}~\bibnamefont {Pitchford}},\ and\ \bibinfo {author} {\bibfnamefont
  {F.}~\bibnamefont {Nori}},\ }\bibfield  {title} {\bibinfo {title} {Qutip 5:
  The quantum toolbox in python},\ }\href
  {https://doi.org/https://doi.org/10.1016/j.physrep.2025.10.001} {\bibfield
  {journal} {\bibinfo  {journal} {Physics Reports}\ }\textbf {\bibinfo {volume}
  {1153}},\ \bibinfo {pages} {1} (\bibinfo {year} {2026})}\BibitemShut
  {NoStop}%
\bibitem [{\citenamefont {Adolphs}\ and\ \citenamefont
  {Renger}(2006)}]{Adolphs2006a}%
  \BibitemOpen
  \bibfield  {author} {\bibinfo {author} {\bibfnamefont {J.}~\bibnamefont
  {Adolphs}}\ and\ \bibinfo {author} {\bibfnamefont {T.}~\bibnamefont
  {Renger}},\ }\bibfield  {title} {\bibinfo {title} {How proteins trigger
  excitation energy transfer in the {{FMO}} complex of green sulfur bacteria},\
  }\href@noop {} {\bibfield  {journal} {\bibinfo  {journal} {Biophys. J.}\
  }\textbf {\bibinfo {volume} {91}},\ \bibinfo {pages} {2778} (\bibinfo {year}
  {2006})}\BibitemShut {NoStop}%
\bibitem [{\citenamefont {Breuer}\ and\ \citenamefont
  {Petruccione}(2002)}]{breuer_theory_2002}%
  \BibitemOpen
  \bibfield  {author} {\bibinfo {author} {\bibfnamefont {H.-P.}\ \bibnamefont
  {Breuer}}\ and\ \bibinfo {author} {\bibfnamefont {F.}~\bibnamefont
  {Petruccione}},\ }\href@noop {} {\emph {\bibinfo {title} {The theory of open
  quantum systems}}}\ (\bibinfo  {publisher} {Oxford University Press on
  Demand},\ \bibinfo {year} {2002})\BibitemShut {NoStop}%
\bibitem [{\citenamefont {Lidar}(2019)}]{lidar2019}%
  \BibitemOpen
  \bibfield  {author} {\bibinfo {author} {\bibfnamefont {D.~A.}\ \bibnamefont
  {Lidar}},\ }\bibfield  {title} {\bibinfo {title} {Lecture {{Notes}} on the
  {{Theory}} of {{Open Quantum Systems}}},\ }\href@noop {} {\  (\bibinfo {year}
  {2019})}\BibitemShut {NoStop}%
\bibitem [{\citenamefont {Mukamel}(1995)}]{MukamelBook}%
  \BibitemOpen
  \bibfield  {author} {\bibinfo {author} {\bibfnamefont {S.}~\bibnamefont
  {Mukamel}},\ }\href@noop {} {\emph {\bibinfo {title} {Principles of Nonlinear
  Spectroscopy}}}\ (\bibinfo  {publisher} {{Oxford University Press}},\
  \bibinfo {address} {{Oxford}},\ \bibinfo {year} {1995})\BibitemShut {NoStop}%
\end{thebibliography}%

\clearpage
\onecolumngrid

\section*{Supplementary Information}
    In the Supplementary Information we provide further details on the computations of the main text. Particularly, we provide: a detailed form of the gFT master equation with each of its terms, details on the correlation and lineshape functions used, 
    information about the symbolic evaluation of the cumulant expressions,
    details on the numerical integration of the integro-differential equation, an analysis on the effects of the phase in the initial coherence and an extensive comparison of gFT with HEOM and non-equilibrium Förster for different values of $\Delta E$.
    
\appendix

\section{Master Equation Terms \label{sec:master_equations}}
Here, we present the derivation of the approximate equation of motion for the reduced density matrix, Eq. (2), of the main text. The general form 
consists of an initial term $\hat{I}(t)$, the Hamiltonian term, and an integral term with a memory kernel. 
The form of these terms can be derived exactly from the Liouville-von Neumann equation for the system and bath state $\hat W(t)$
\begin{eqnarray}
    \frac{\partial}{\partial t}\hat W(t)=-\frac{i}{\hbar}[\hat H,\hat W(t)].
\end{eqnarray}
The Hamiltonian is split into the unperturbed part, $\hat{H}_0$, and the interaction part, $\hat{V}$. In the Förster case, the operator $\hat{V}$ does not depend on the bath degrees of freedom. Using the interaction picture with respect to $\hat H_0$, we obtain
\begin{eqnarray}
\label{eq:LvN_int}
    \frac{\partial }{\partial t}\hat W^{(I)}(t)=-\frac{i}{\hbar}[\hat{V}(t),\hat W^{(I)}(t)],
\end{eqnarray}
where $\hat W^{(I)}(t)=\hat U_0^{\dagger}(t)\hat W(t)\hat U_0(t)$ and $\hat{V}(t)=\hat U_0^{\dagger}(t)\hat{V}\hat U_0(t)$. By integrating Eq. (\ref{eq:LvN_int}) and reinserting the result into the same equation, we obtain
\begin{eqnarray}
    \frac{\partial }{\partial t}\hat W^{(I)}(t)=-\frac{i}{\hbar}[\hat{V}(t),\hat W^{(I)}(0)]-\frac{1}{\hbar^2}\int\limits_{0}^{t}{\rm d}\tau [\hat{V}(t),[\hat{V}(\tau),\hat W^{(I)}(\tau)]].
\end{eqnarray}
Returning to the Schrödinger picture, $\frac{\partial }{\partial t}\hat W(t) = - \frac{i}{\hbar} [\hat{H}_0, \hat{W}(t)] + \hat{U}_0(t)\frac{\partial \hat W^{(I)}(t)}{\partial t} U_0^{\dagger}(t)$, and performing a substitution $\tau^{\prime}=t-\tau$ we obtain
\begin{eqnarray}
\label{eq:final_identity}
    \frac{\partial }{\partial t}\hat W(t)=-\frac{i}{\hbar}[\hat{V},\hat U_0(t)\hat W(0)\hat U_0^{\dagger}(t)]-\frac{i}{\hbar}[\hat H_0,\hat W(t)]-\frac{1}{\hbar^2}\int\limits_{0}^{t}{\rm d}\tau' [\hat{V},\hat U_0(\tau^{\prime})[\hat{V},\hat W(t-\tau^{\prime})]\hat U_0^{\dagger}(\tau^{\prime})].
\end{eqnarray}
This is Eq. (\ref{eq:general_eom_w}) in the main text. 

By using a formally exact factorization of the density matrix elements 
\begin{eqnarray}
    \hat{W}(t)=\sum_{a,b} \tr_B\{\hat W_{ab}(t)\} \ket{a}\bra{b}\:  \:\frac{\hat W_{ab}(t)}{\tr_B\{\hat W_{ab}(t)\}}=\sum_{a, b}\rho_{ab}(t) \ket{a}\bra{b} \:  \:\hat{w}_{ab}(t),
\end{eqnarray}
and assuming zero-order approximation for the bath relative operators
\begin{eqnarray}
\label{eq:state_of_bath}
    \hat{w}_{ab}(t)=\frac{\hat{u}_{a}(t)\hat{w}_{eq}\hat{u}_{b}^{\dagger}(t)}{{\tr}_B\{\hat{u}_{a}(t)\hat{w}_{eq}\hat{u}_{b}^{\dagger}(t)\}}=\frac{\hat{u}_{a}(t)\hat{w}_{eq}\hat{u}_{b}^{\dagger}(t)}{G_{ab}(t)},
\end{eqnarray}
we can formally trace over the bath degrees of freedom in Eq. (\ref{eq:final_identity}) and obtain the term of the reduced density matrix master equation, as shown below. Note that contrary to the standard approach in Open Quantum Systems theory \cite{breuer_theory_2002} we do not need to assume a factorized initial state of system and bath.

\subsection{Initial Term}
For the initial condition term (the first term on the r.h.s. of Eq. (\ref{eq:equation_foerster})), we obtain 
\begin{eqnarray}
    \hat{I}(t)&=&-\frac{i}{\hbar}{\rm tr}_{B}\{[\hat{V},U_0(t)\rho(0)\hat{w}_{eq}U_0^{\dagger}(t)]\}=-\frac{i}{\hbar}\sum_{ab}[\hat{V},|a\rangle D_{ab}(t)\rho_{ab}(0)\langle b|]
    \\
   &=&   -\frac{i}{\hbar}\sum_{ab}  \rho_{ab}(0) D_{ab}(t)\sum_c \Big(J_{ca}\ket{c}\bra{b} - J_{bc}\ket{a}\bra{c}\Big), 
\end{eqnarray}
where we introduced
\begin{equation}
\begin{split}
\label{eq:G_ab} 
    D_{ab}(t)&={\tr}_{B}\{\hat u_{a}(t)\hat{w}_{eq}\hat u_{b}^{\dagger}(t)\} e^{- i \omega_{ab}t} \\
    &=  e^{-g_{aa}(t)-g^{*}_{bb}(t)+2{\rm Re} [g_{ab}(t)]} e^{- i \omega_{ab}t}
    \end{split}
\end{equation}
and used the cumulant expansion to evaluate it in the second line. The lineshape function $g(t)$ is detailed in Eq.~(\ref{eq:lineshape}). 
The initial term elements $I_{ab}(t)$ thus read
\begin{eqnarray}
    I_{ab}(t)=\langle a|\hat{I}(t)|b \rangle = -\frac{i}{\hbar}\left(J_{ac}D_{cb}(t)\rho_{cb}(0)-D_{ac}(t)\rho_{ac}(0)J_{cb}\right).
\end{eqnarray}

The initial term can always be removed by rescaling the bath operators if the bath state is time-independent \cite{lidar2019}, note however, that we assume a non-equilibrium state of the bath.

\subsection{Hamiltonian and Dephasing Terms}

For the Hamiltonian term of Eq. (\ref{eq:equation_foerster}) (the second term on the r.h.s.), 
we obtain
\begin{eqnarray}\label{eq:Lrho}
    \mathcal L(t)\hat \rho(t)=\frac{1}{\hbar}[H_S,\rho(t)]-i\mathcal D(t)\hat{\rho}(t)
    \: =\hat{\Omega}(t) \circ \hat{\rho}(t).
\end{eqnarray}
The second term represents the dephasing due to the bath evolution, 
\begin{eqnarray}\label{eq:mathcalD}
    \mathcal D(t)\hat{\rho}(t)=-\frac{i}{\hbar}\sum_{ab} \left(x_{\underline{a}b}(t) - x_{a\underline{b}}(t)\right)  \rho_{ab}(t)|a\rangle\langle b|,
\end{eqnarray}
which depends on the shift of the environment due to the electronic excitation, captured by the function
\begin{equation}\label{eq:x}
x_{\underline{a}b}(t) \equiv \Tr_B\left( \hat{X}_a \hat{w}_{ab}(t)\right).
\end{equation}
In this notation, we highlight the state that experiences the electronic transition with the underlined subscript---so $x_{a\underline{b}}(t) =\Tr_B\big(  \hat{w}_{ab}(t) \hat{X}_b\big) = x_{\underline{b}a}^*(t)$. 
We need to evaluate 
\begin{eqnarray}
\label{eq:x_cum}
    \Tr_B\{\hat{u}_b^{\dagger}(t)\hat{X}_a \hat{u}_a(t)w_{eq}\} = -i(\dot{g}_{aa}(t)-\dot{g}_{ab}^{*}(t))e^{-g_{aa}(t)-g^{*}_{bb}(t)+2{\rm Re}g_{ab}(t)}
\end{eqnarray} which gives, after dividing by the normalization $G_{ab}(t)$ 
%
%
\begin{equation}
    x_{\underline{a}b}(t)= -i\dot{g}_{aa}(t)+i\dot{g}_{ab}^{*}(t),
\end{equation}
and
\begin{equation}
    x_{a\underline{b}}(t)=i\dot{g}_{bb}^{*}(t)-i\dot{g}_{ab}(t). 
\end{equation}

In the equation of motion, these terms appear as a difference, cf. Eq. (\ref{eq:Lrho}), which can be expressed by the frequency
\begin{eqnarray}
\label{eq:cum_Ome_ab}
    \Omega_{ab}(t) &\equiv& \omega_{ab} + \Tr_B(\hat{X}_a \hat{w}_{ab}(t) ) - \Tr_B( \hat{w}_{ab}(t) \hat{X}_b)\\
    &=&\omega_{ab} +  x_{\underline{a}b}(t) - x_{a\underline{b}}(t)  \nonumber ,
\end{eqnarray}
written using the lineshape function $g(t)$ in Eq. (8) of the main text. 
Note that, by construction, $\Omega_{aa}(t)=0$.


%
\subsection{Memory kernel}
\begin{figure}
    \includegraphics[]{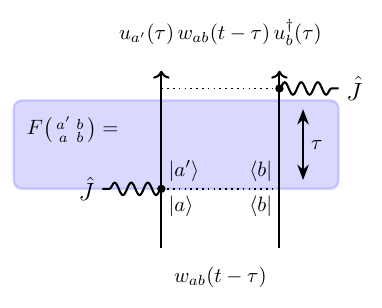}
    \caption{Illustration of the bath evolution correlators, Eq. \ref{eq:f_box}.
    \label{fig:f_box}}
\end{figure}

The integral kernel of the third term on the r.h.s. of Eq. (\ref{eq:equation_foerster}) leads to the superoperator
\begin{eqnarray}
\label{eq:M_term}
    {\cal M}(t,\tau)\rho(t-\tau)= 
        \sum_{ab}[
            \hat{V},
            \tr_B\Big\{
                \hat{U}_0(\tau)[\hat{V},|a\rangle\langle b|]\hat{w}_{ab}(t-\tau)\hat{U}_0^{\dagger}(\tau)
                \Big\}
            ]\rho_{ab}(t-\tau) \ket{a}\bra{b}.
\end{eqnarray}
To express this equation, we define the 
bath evolution correlators 
\begin{eqnarray}
\label{eq:f_box}
    F_{ab}^{a'b} (t,\tau)\equiv \tr_{B}\{\hat{u}_{a'}(\tau)\hat{w}_{ab}(t-\tau)\hat{u}_{b}^{\dagger}(\tau)\}e^{-i\omega_{a'b}\tau}, \\
    F_{ab}^{a'b} (t,\tau)e^{i\omega_{a'b}\tau} \approx f_{ab}^{a'b} (t,\tau) \equiv \tr_{B}\{\hat{u}_{a'}(\tau)\hat{u}_a(t-\tau)\hat{w}_{eq}\hat{u}_{b}^{\dagger}(t-\tau)\}\hat{u}_{b}^{\dagger}(\tau)\},
\end{eqnarray}
where the box notation is inspired by the Feynman double-sided diagram, as illustrated in Fig. \ref{fig:f_box}. The explicit form for this function is  given in Eq. (\ref{eq:fcda_cummulant}) of the main text.

We further introduce the system operators $\hat{V}_{\underline{a}b}(t,\tau)$ and $\hat{V}_{a\underline{b}}(t,\tau)$
that describe the free evolution of the relative bath state $\hat{w}_{ab}$ from time $t-\tau$ to $t$ 
after one interaction $V$, that is, allowing for one electronic transition on the state which index is underlined. 
 Namely,


\begin{eqnarray}
    \hat{V}_{\underline{a}b}(t,\tau) 
    &=& {\tr}_B\{  \hat{U}_0(\tau)\hat{V}|a\rangle \hat{w}_{ab}(t-\tau)\langle b| \hat{U}_0^{\dagger}(\tau)\}\nonumber \\
    &=&\sum_{a'}J_{a'a} e^{-\frac{i}{\hbar}\varepsilon_{a'} \tau} \:  f^{a'b}_{ab}(t,\tau) \: e^{\frac{i}{\hbar}\varepsilon_b \tau}|a'\rangle\langle b|,
\end{eqnarray}
\begin{eqnarray}
    \hat{V}_{a\underline{b}}(t,\tau)
    &= \tr_B\{U_0(\tau)|a\rangle w_{ab}(t-\tau)\langle b|\hat{V}U_0^{\dagger}(\tau)\} \nonumber \\
    &=\sum_{b'}J_{ab'} e^{-\frac{i}{\hbar}\varepsilon_{a} \tau} \:  f^{ab'}_{ab}(t,\tau) \: e^{\frac{i}{\hbar}\varepsilon_{b'} \tau}|a\rangle\langle b'|.
\end{eqnarray}

The memory kernel has four contributions corresponding to the two nested commutators
%
\begin{eqnarray}
   \mathcal{M}^{ab}_{cd}(t,\tau)= \langle a|[\hat{V},\hat{V}_{\underline{c}d}(t,\tau)-\hat{V}_{c\underline{d}}(t,\tau)]|b\rangle \nonumber \\  
  = \bra{a}   \hat{V}  \hat{V}_{\underline{c}d}(t,\tau) \ket{b}   - \bra{a}\hat{V}_{\underline{c}d}(t,\tau)\hat{V}\ket{b} 
  -\langle a|\hat{V}\hat{V}_{c\underline{d}}(t,\tau)|b\rangle+\langle a|\hat{V}_{c\underline{d}}(t,\tau)\hat{V}|b\rangle.
\end{eqnarray}
When the operator $\hat{V}_{\underline{c}d}$ acts on the vector $|b\rangle$, the only value of $d$ that leads to non-zero results is $d=b$, so that $\hat{V}_{\underline{c}d}(t,\tau)|b\rangle= \delta_{db} \hat{V}_{\underline{c}b}(t,\tau)|b\rangle$, similarly in the expression $\langle a|\hat{V}_{c\underline{d}}(t,\tau)=\delta_{ac}\langle a|\hat{V}_{a\underline{d}}(t,\tau)$. Correspondingly, we have
\begin{eqnarray}
   &&
   {\cal M}^{ab}_{cd}(t,\tau)=  \langle a|\hat{V}\hat{V}_{\underline{c}b}(t,\tau)|b\rangle\delta_{bd} +\langle a|\hat{V}_{a\underline{d}}(t,\tau)\hat{V}|b\rangle\delta_{ac} - \langle a|\hat{V}_{\underline{c}d}(t,\tau)\hat{V}|b\rangle -\langle a|\hat{V}\hat{V}_{c\underline{d}}(t,\tau)|b\rangle.
\end{eqnarray}
Using eqs. (\ref{eq:G_ab}), (\ref{eq:cum_Ome_ab}) and (\ref{eq:f_box}) we write a closed set of equations for all elements of the RDM $\rho_{ab}(t)$.


%
%
%
%

\subsection{Full equation of motion for the RDM}
The operators and superoperators defined above describe the evolution of the reduced density matrix $\hat{\rho}(t)$ exactly. The corresponding RDM equation
\begin{eqnarray}
 \frac{\partial }{\partial t}\hat{\rho}(t)=\hat{I}(t)-\mathcal D(t)\hat{\rho}(t)-\frac{i}{\hbar}[\hat H_S,\hat \rho(t)]-\sum_{cd}\int\limits_{0}^{t}{\rm d}\tau \Big[\hat{V},             
 \big(\hat{V}_{{\underline{c}}d}(t,\tau) - \hat{V}_{c{\underline{d}}}(t,\tau) \big)
 \Big]
 \rho_{cd}(t-\tau). 
\end{eqnarray}
This is an exact equation of motion for the RDM of an open quantum system with discrete levels. Its usage is only hindered by the need to evaluate Eqs. (\ref{eq:G_ab}), (\ref{eq:x}) and (\ref{eq:f_box}) which cannot be in general done exactly. Represented in the basis of eigenstates $|a\rangle$ of the Hamiltonian $H_S$ we get
\begin{eqnarray}
\label{eq:full_eq_in_basis}
    \frac{\partial }{\partial t}\rho_{ab}(t)=I_{ab}(t)-i\Omega_{ab}(t)\rho_{ab}(t)-\sum_{cd}\int\limits_{0}^{t}{\rm d}\tau {\cal M}^{ab}_{cd}(t,\tau) \rho_{cd}(t-\tau), 
\end{eqnarray}
where $\omega_{ab}=\frac{\varepsilon_{a}-\varepsilon_b}{\hbar}$ are the transition frequencies between the eigenstates and all other terms are explicitly defined above.

\subsection{Relaxation Rates}

The particularly important case of the integral term is the contribution by which a diagonal element of the density matrix feeds into the change of another diagonal element. An integral term involving an integrand factor $\rho_{ab}(t-\tau)$ can be trivially time-localized by assuming $\rho_{ab}(t-\tau)\approx\rho_{ab}(t)$.
We obtain time-dependent relaxation rates between electronic states:
\begin{eqnarray}
    K_{ab}(t)=2{\rm Re}\int\limits_{0}^{t}{\rm d}\tau {\cal M}^{aa}_{bb}(t,\tau)=\frac{2|J_{ab}|^2}{\hbar^2}{\rm Re}\int\limits_{0}^{t}{\rm d}\tau\; e^{-g_{aa}(\tau)-g_{bb}(\tau)+2{\rm Im}[g_{bb}(t-\tau)-g_{bb}(t)]-i\omega_{ab}\tau}
\end{eqnarray}
These rates become the standard Förster resonance energy transfer rates upon the limit $t\rightarrow\infty$.

\section{Bath correlation functions and lineshape functions \label{sec:bath_corfce}}

In this work, we assume a very simple model of the bath characterized by the Drude-Lorentz spectral density
\begin{eqnarray}
    J(\omega)= \dfrac{2 \lambda \gamma \omega}{\omega^2 + \gamma^2}
\end{eqnarray}
with bandwidth $\gamma$ and interaction strength $\lambda$. This leads to the two-point energy gap correlation function in the form \cite{MukamelBook}
\begin{eqnarray}
  &&  C_{aa}(t)={\rm tr}_{B}\{U^{\dagger}_g(t)\hat{X}_{a}U_g(t)\hat{X}_a \hat{w}_{eq}\}=  \nonumber \\
  &&  =\frac{\lambda_a}{\tau_a}\Big(\cot\!\frac{\beta\hbar}{2\tau_a}-i\Big)e^{-t/\tau_a}
\;+\;\frac{4\lambda_a}{\beta\hbar\,\tau_a}\sum_{k=1}^{\infty}
\frac{\nu_k}{\nu_k^{2}-\tau_a^{-2}}\,e^{-\nu_k t},
\qquad t\ge 0,
\end{eqnarray}
where $\lambda_a$ is the bath reorganization energy at site $a$ and $\tau_{a}=\frac{1}{\gamma_a}$ the bath correlation time at site $a$. 
The corresponding lineshape function and its derivative read
\begin{eqnarray}\label{eq:lineshape}
    g_{aa}(t)=\frac{1}{\hbar^2}\int\limits_{0}^{t}{\rm d}\tau \int\limits_{0}^{\tau}{\rm d}\tau^{\prime}C_{aa}(\tau^{\prime}) =  \frac{2}{\pi \hbar} \int_0^t d \tau \int_0^\tau d \tau' \int_{-\infty}^\infty d \omega J_a(\omega) (\coth \frac{\beta \hbar \omega}{2} \cos (\omega \tau') - i \sin (\omega \tau')),  
\end{eqnarray}
\begin{eqnarray}
    \dot{g}_{aa}(t)=\frac{1}{\hbar^2}\int\limits_{0}^{t}{\rm d}\tau\;C_{aa}(\tau).
\end{eqnarray}
In numerical simulations, the correlation functions and the lineshape functions are obtained by applying the Pad\`e expansion to the temperature-dependent part so that the whole correlation function can be approximated by $C_{aa}(\tau) \approx \sum_{k=0}^{K} c_{a,k} e^{-\gamma_k \tau}$. In our computations, we truncate at $K=1$ as this proves sufficient given the smoothness of the Drude-Lorentz spectral density, in the numerical results this is computed using the $\texttt{DrudeLorenzPadeBath()}$ function of QuTiP \cite{lambert_qutipv5_2026} with $\texttt{Nk}=K=1$.

\section{Evaluation of cumulants \label{sec:cumulants}}

Cumulant expressions are evaluated using the symbolic module of the Quantarhei package \cite{Mancal2025_Quantarhei} (tested with version 0.0.68) using the technique described, e.g. in \cite{MukamelBook}. Expressions with bath evolution operators $u_a(t)$ relative to electronic states $|a\rangle$ have to be rearranged into pairs together with the counter-propagating ground state evolution operator $u_g(t)$ and rewritten into the Quantarhei notation where
\begin{eqnarray}
    u_a(t)u^{\dagger}_g(t) \rightarrow {\rm Uegd(a,t)}, \\
    u^{\dagger}_a(t)u_g(t) \rightarrow {\rm Uedg(a,t)}, \\
    u_g(t)u^{\dagger}_a(t) \rightarrow {\rm Uged(a,t)}, \\
    u^{\dagger}_g(t)u_a(t) \rightarrow {\rm Ugde(a,t)}.
\end{eqnarray}
The pairs of evolution operators are preresented by $\rm Uxyz(i,t)$ functions, where the letter code is the following: $\rm g$ at the $\rm xyz$ positions represents the ground state evolution $u_g(t)$, $\rm gd$ represents $u^{\dagger}_g(t)$, $\rm e$ represents any of the excited state evolutions, such as $u_a(t)$, and $ed$ represents $u^{\dagger}_a(t)$. The index $\rm a$ of the excited state becomes the index $\rm i$ in the argument of the $\rm Uxyz$ function. The last argument is the time.

The approximate correlator of Eq. (\ref{eq:f_box}) (note that $b'=b$ and $a'=c$ here) reads in this notation as follows
\begin{eqnarray}
\label{eq:f_b_pairs}
    f_{ab}^{cb} (t,\tau)\equiv \tr_{B}\{\hat{u}_{c}(\tau)u_a(t-\tau)\hat{w}_{eq}u^{\dagger}_{b}(t)\} = \tr_{B}\{u^{\dagger}_{b}(t)\hat{u}_{c}(\tau)u_a(t-\tau)\hat{w}_{eq}\} = \nonumber \\
    = \tr_{B}\{u^{\dagger}_{b}(t)u_g(t)u_g^{\dagger}(t)u_c(t)u^{\dagger}_c(t-\tau)u_g(t-\tau)u^{\dagger}_g(t-\tau)u_a(t-\tau)\}.
\end{eqnarray}
The expression in trace in Eq. (\ref{eq:f_b_pairs}) is translated as
\begin{eqnarray}
    u^{\dagger}_{b}(t)u_g(t)u_g^{\dagger}(t)u_c(t)u^{\dagger}_c(t-\tau)u_g(t-\tau)u^{\dagger}_g(t-\tau)u_a(t-\tau) \rightarrow \nonumber \\
    \rightarrow {\rm Uedg(b,t)*Ugde(c,t)*Uedg(c,t-tau)*Ugde(a,t-tau)}
\end{eqnarray}
This expression is evaluated by Quantarhei symbolically as
\begin{eqnarray}
    {\rm -conjugate(gg(b, b, t)) + conjugate(gg(b, b, t - tau)) + conjugate(gg(b, c, t)) } \nonumber \\
    {\rm - conjugate(gg(b, c, t - tau)) + gg(b, a, t) - gg(b, a, tau) - gg(b, a, t - tau) + gg(b, c, tau) - gg(c, a, t) } \nonumber \\
    {\rm + gg(c, a, tau) + gg(c, a, t - tau) - gg(c, c, tau)}
\end{eqnarray}
which translates into
\begin{eqnarray}
\label{eq:f_in_g}
    e^{-g^{*}_{bb}(t) + g^{*}_{bb}(t - \tau) + g^{*}_{bc}(t) - g^{*}_{bc}(t - \tau) + g_{ba}(t) - g_{ba}(\tau)}  \nonumber \\
    \times e^{- g_{ba}(t - \tau) + g_{bc}(\tau) - g_{ca}(t) + g_{ca}(\tau) + g_{ca}(t - \tau) - g_{cc}(\tau)}.
\end{eqnarray}
The set of lineshape functions, Eq. (\ref{eq:f_in_g}) corresponds to Eq. (\ref{eq:fcda_cummulant}) of the main manuscript. Note that Quantarhei returns the expected argument of the exponential cumulant expression.

Similarly, we can evaluate more complicated cumulants, such as Eq. (\ref{eq:x}), where 
\begin{eqnarray}
    \Tr_B\{\hat{u}_b^{\dagger}(t)\hat{X}_a \hat{u}_a(t)w_{eq}\} = 
    \Tr_B\{\hat{u}_g(t)\hat{u}_b^{\dagger}(t)\hat{X}_a \hat{u}_a(t)\hat{u}_g^{\dagger}(t)w_{eq}\}=\frac{\partial}{\partial x}\Tr_B\{\hat{u}_g(t)\hat{u}_b^{\dagger}(t)e^{x\hat{X}_a\tau} \hat{u}_a(t)\hat{u}_g^{\dagger}(t)w_{eq}\}\Big|_{x=0;\tau=0}
\end{eqnarray}
translates into
 \begin{eqnarray}
    {\rm Uged(b,t)*ExpdV(a,tau,x)*Uegd(a,t)},
\end{eqnarray}
in which we set $x=0$ and $\tau=0$ after the evaluation.

All cumulants required in this work can be evaluated by the Python script, \verb|cumulants.py|, attached to this SI.   

\section{Numerical integration of integro-differential equation \label{sec:numerics}}

To calculate the evolution of the system, we treatEq. (\ref{eq:equation_foerster}) of the main text numerically by step-wise integrating both sides with a time step $\Delta t$.
Using the Euler method, we arrive at an explicit expression for the evolution.
Rearranged by common factors of $\rho(\cdot)$ we find

\begin{align} \label{app:numerical_integration}
\begin{split}
    \rho (t + \Delta t) - \rho(t) = \text{ } &\rho(0) \left[ \dfrac{I(t + \Delta t) + I(t)}{2} \Delta t + \dfrac{\mathcal{M}(t,0) + \mathcal{M}(t+\Delta t,0)}{4} \Delta t^2 \right] + \\
                                  + &\rho(t) \left[ \dfrac{\mathcal{L}(t)}{2} \Delta t + \dfrac{\mathcal{M}(t+\Delta t,t)}{2} \Delta t + \dfrac{\mathcal{M}(t,t)}{4} \Delta t^2 \right] + \\
                                  + &\rho(t+\Delta t) \left[ \dfrac{\mathcal{L}(t + \Delta t)}{2} \Delta t + \dfrac{\mathcal{M}(t + \Delta t, t + \Delta t)}{4} \Delta t^2 \right] + \\
                                  + &\sum_{i=1}^{N_t -1} \rho(i\Delta t) \left[ \dfrac{\mathcal{M}(t, i \Delta t) + \mathcal{M}(t + \Delta t, i \Delta t)}{2} \right] \Delta t^2.
\end{split}
\end{align}

Here, $N_t = (t + \Delta t) / \Delta t$ refers to the number of time steps from $0$ to the new time $t + \Delta t$.
The stability of the integration method is guaranteed by comparison with the results from HEOM.

%


%

\section{The Influence of Time Non-Local Effects}
\begin{figure}[ht]
    \centering
    \includegraphics[width=\linewidth]{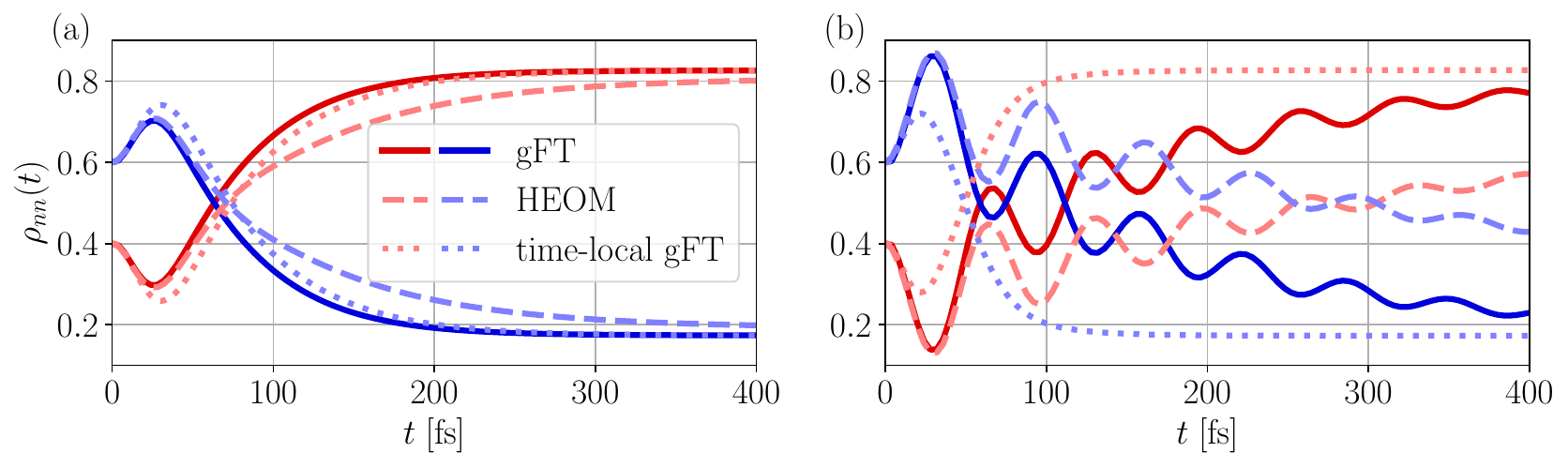}
    \caption{Population evolutions comparing gFT, HEOM and the time localized version of gFT. For all calculations, $\Delta E =~300\; \mr{cm}^{-1}$, $t_c = 30\; \mr{fs}$ and $T = 277\; \mr{cm}^{-1}$. In (a) $J=110\; \mr{cm}^{-1}$, $\lambda=175\; \mr{cm}^{-1}$. In (b) $J=200\; \mr{cm}^{-1}$, $\lambda=25\; \mr{cm}^{-1}$.}
    \label{fig:time_local_nonlocal}
\end{figure}

We study two versions of the gFT theory, namely, the full time non-local one, and the time-localized version where, for the population of the state $|a\rangle$, we replace $\rho_{aa}(t-\tau)$ with $\rho_{aa}(t)$.
We compare two example dynamics in Fig. \ref{fig:time_local_nonlocal}. In Fig. \ref{fig:time_local_nonlocal}a, where $J<\lambda$ we notice that the initial population dynamics is much better captured by the time non-local gFT, while at long times the two versions of the gFT theory converge to each other, both deviating similarly from the results of HEOM.
With larger coupling $J>\lambda$, HEOM and gFT predict oscillatory dynamics and a much slower equilibrium time than time-localized gFT.
This can be seen in Fig. \ref{fig:time_local_nonlocal}b.
The time localized theory deviates from HEOM much earlier than the full time non-local theory. We can also see that the lasting oscillations on the population dynamics are only captured by the time non-local theory. 

\section{Comparison with Numerically Exact Propagation\label{sec:additional_data}}

In this section, we complement the parameter values shown in the main text with additional values of $\Delta E$ and more examples of evolutions.
Due to the restriction to the single excitation of the dimer, the density matrix of the dimer is effectively that of a two level system and can thus be represented in the Bloch representation $\ha \rho = \frac{1}{2}(\hat{\openone} + x \hat \sigma_x + y \hat \sigma_y + z \hat \sigma_z)$, where the coordinates $x,\, y$ encode the real and imaginary part of the coherence, respectively, and $z$ encodes the population imbalance between the two sites. The trace distance between two density matrices has a very simple representation in the Bloch coordinates
\begin{equation}
    T(\hat \rho, \hat \rho') = \sqrt{(x-x')^2 + (y-y')^2 + (z-z')^2}.
\end{equation}

The error characterizes the maximum distance between two evolutions.
One of the evolutions will be the reference numerical exact dynamics generated by HEOM, and the other evolution will be either the generalized F\"orster theory (gFT) developed in this work or the non-equilibrium version of the F\"orster theory (non-eq. FT).
We consider errors for the populations and the coherences individually by splitting the contributions of the trace distance into the population part $\varepsilon_{pop}$ and the coherence part $\varepsilon_{coh}$.
With this we aim to better highlight the accuracy of the presented formalism.
Since gFT cannot reproduce steady state delocalization due to the restriction to the site basis, we expect the error in populations to be much less than that of the coherences.
We also differentiate between short and intermediate times.
This leads us to define our error measures as

\begin{subequations}
\begin{align}
    \varepsilon_{pop}^{\rm short} &= \max_{t \in [0, t_c]} |z^{\rm HEOM}(t) - z^{\rm gFT}(t)|,\\
    \varepsilon_{pop}^{\rm inter} &= \max_{t \in [t_c, t_s]} |z^{\rm HEOM}(t) - z^{\rm gFT}(t)|, \\
    \varepsilon_{coh}^{\rm short} &= \max_{t \in [0, t_c]} \sqrt{(x^{\rm HEOM}(t) - x^{\rm gFT}(t))^2+(y^{\rm HEOM}(t) - y^{\rm gFT}(t))^2},\\
    \varepsilon_{coh}^{\rm inter} &= \max_{t \in [t_c, t_s]} \sqrt{(x^{\rm HEOM}(t) - x^{\rm gFT}(t))^2+(y^{\rm HEOM}(t) - y^{\rm gFT}(t))^2}.
    \end{align}
    \end{subequations}
We hereby use the correlation time $t_c$ to define the end of \emph{short} times.
The \emph{intermediate} time is then taken from $t_c$ until the steady state time $t_s$.
This time is determined by setting a threshold on the rate of change of populations at advanced times (in cases where no steady state is reached, the simulation time is taken instead).

For consistency throughout the main text and in the SI we restrict ourselves to displaying the regimes of low errors between $\varepsilon \in [0, 0.1]$, since those highlight where the theories are valid.
We consider $\varepsilon>0.1$ as cases where the theory does not closely describe the real dynamics and we therefore do not discuss them in detail.

\subsection{Molecules on resonance $\Delta E =0$}

\begin{figure}[ht!]
    \centering
    \includegraphics[width=\linewidth]{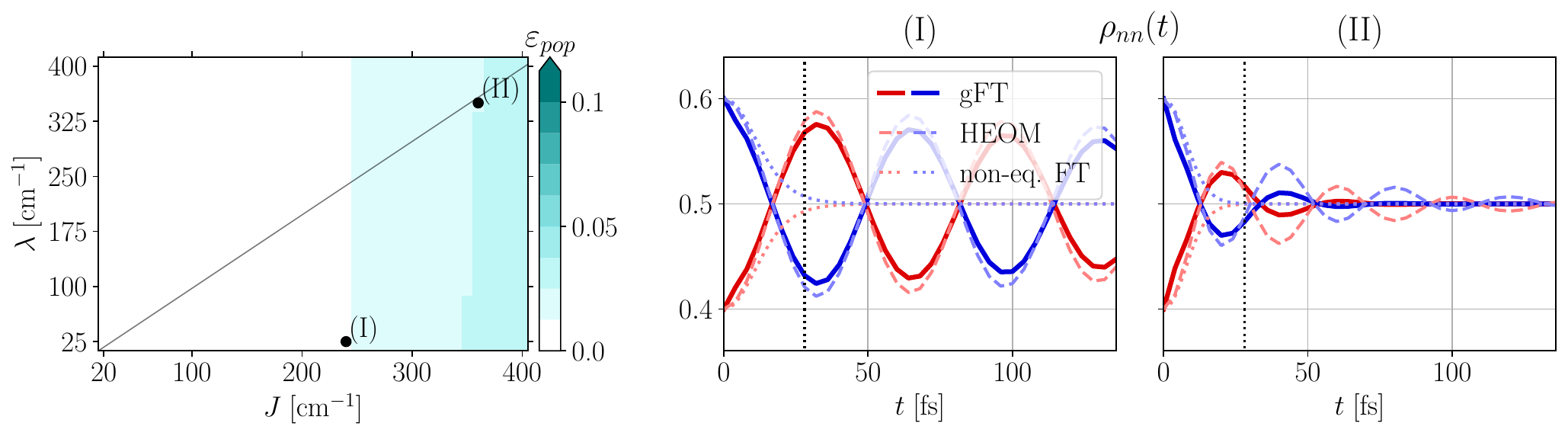}
    
    \caption{Population-only trace distance comparing gFT and HEOM at short times and $\Delta E = 0$ with example evolutions showing the whole population dynamics.}
    \label{fig:short-pop-de0}
\end{figure}

\begin{figure}[ht!]
    \centering
    \includegraphics[width=\linewidth]{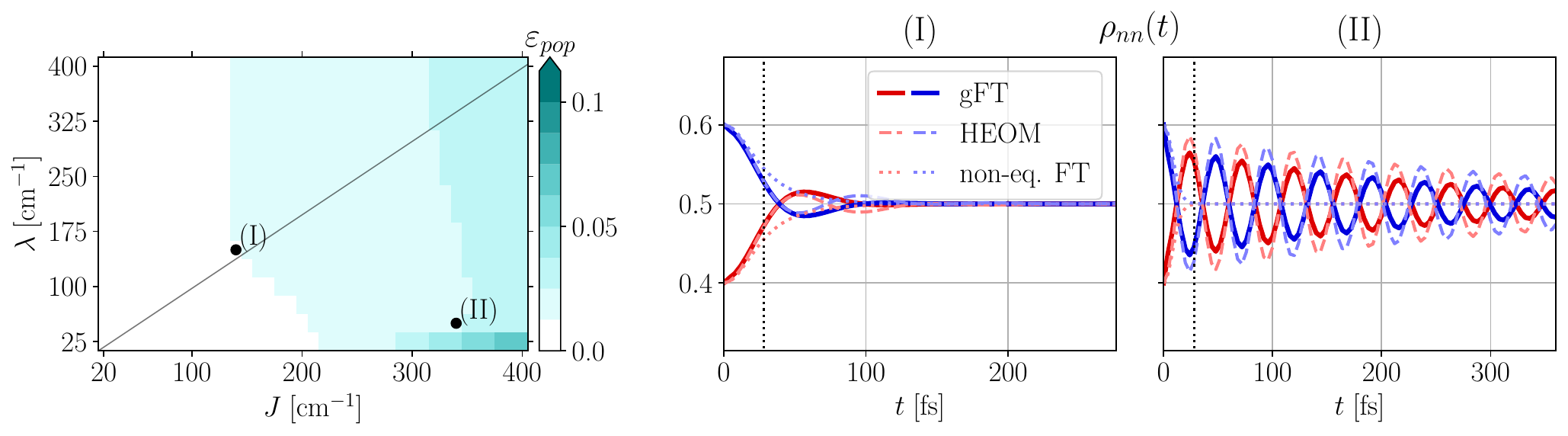}

    \caption{Population-only trace distance comparing gFT and HEOM at intermediate times and $\Delta E = 0$ with example evolutions showing the whole population dynamics. Although hard to see, the dotted line non-eq. FT result quickly converges to $0.5$.}
    \label{fig:inter-pop-de0}
\end{figure}

Figures \ref{fig:short-pop-de0} and \ref{fig:inter-pop-de0} show the error phase diagram as well as example evolutions of population dynamics in the degenerate case $\Delta E=0$.
The first significant point of these evolutions is that the two populations will converge to the equilibrium value given by a uniform distribution between the two levels $\lim_{t\to \infty }\rho_{ii}=\frac{1}{2}$.
The phase diagram in Fig. \ref{fig:short-pop-de0} shows that the short time evolution modeled by gFT follows that of HEOM very closely.
Indeed, the maximum value of the error in the phase diagram is $\max_{\lambda,J}\varepsilon_{pop}<0.04$.
We also see that while the error depends on the hopping $J$, it does not depend on the reorganization energy $\lambda$.
The reason for this is clear, the gFT predicts very correct behavior in short time intervals, especially due to the presence of the inhomogeneous term $\hat I(t)$.
In particular, this makes the short time error show barely any dependence on $\lambda$.

In evolution (I), we observe oscillatory behavior for weak system bath coupling $\lambda$ and moderate hopping $J$.
This evolution shows many oscillations, and we do not see the times for which it stabilizes to the steady state.
Note that for this case, non-eq. FT (dotted lines) predicts a very fast decay to the equilibrium value.
This shows that our gFT works at short times in a regime where the Förster theory was not even expected to work (weak system-bath coupling).

Evolution (II) shows the evolution for strong system-bath coupling $\lambda$ and strong hopping $J$.
This is another example of a case where Förster theories are not guaranteed to work.
However, we see that at short times gFT closely reproduces the HEOM damped oscillations, with an error slightly higher than the one in the previous evolution.
In comparison, non-eq. FT again fails to model these oscillations and predicts monotonous convergence to the steady state.
For longer times, we see that the gFT slightly underestimates the following bumps in the oscillation. 

Figure \ref{fig:inter-pop-de0} shows that gFT also provides a good model for intermediate times. In the phase diagram we observe that $\varepsilon_{pop}$ remains quite small $\varepsilon \approx 0.03$.
Comparing with the short time error phase diagram in the previous figure we now see a more nontrivial dependence with the parameters.
In particular, for smaller values of $J$ we have an error $\varepsilon_{pop} < 0.02$.
Interestingly, the boundary of this region does not depend on $\lambda$ for $\lambda>J$.
For $\lambda<J$ we see that the error increases with a stronger system-bath coupling.
This is quite counter-intuitive, as the opposite (i.e. larger error at stronger system-bath coupling) is expected.
Evolution (I) shows an example dynamics where gFT closely reproduces the damped oscillations of HEOM, while non-eq. FT predicts monotonous decay.
Evolution (II), in the weak system bath coupling regime, shows that gFT models the HEOM evolution with reasonable accuracy, even out of its regime of applicability for intermediate times.
In particular we see that the dynamics has some fast oscillations that slowly decay towards equilibrium.
gFT also predicts those oscillations, only with slightly smaller peaks.
Close inspection shows that non-equ. FT theory (dotted line) completely misses the damped oscillations.

\begin{figure*}[ht]
    \centering
    \includegraphics[width=\linewidth]{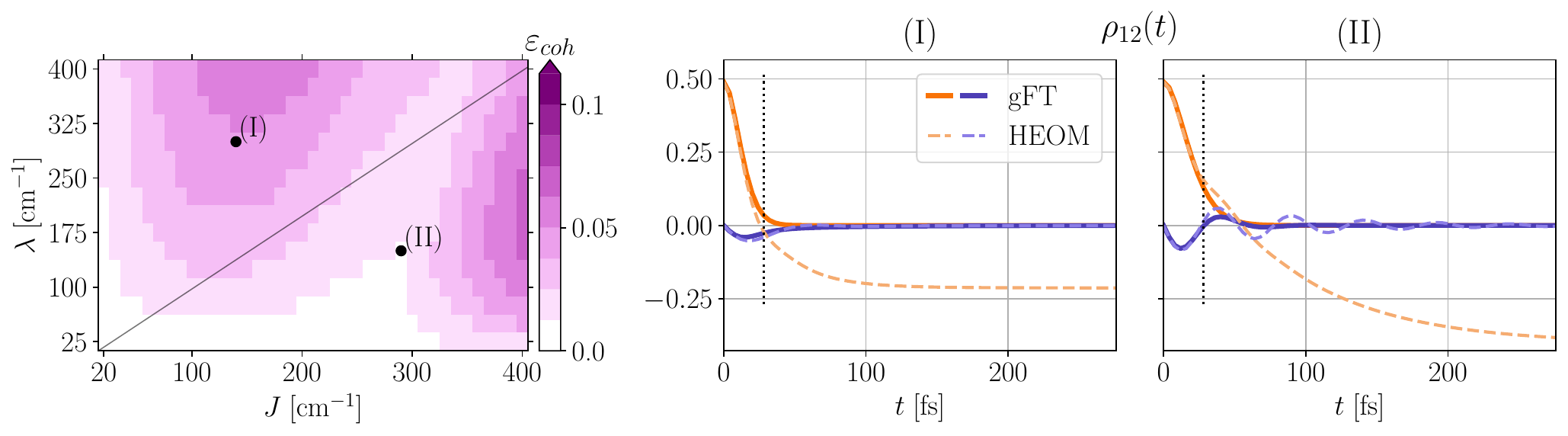}
    
    \caption{Phase diagram (left) for the short time difference of coherences $\varepsilon_{coh}$ between generalized Förster Theory and HEOM for degenerate levels $\Delta E = 0\, \mr{cm}^{-1}$. Plots (I) and (II) depict two example time evolution comparing gFT (solid) and HEOM (dashed). Evolution (I) is taken in the blob at strong system bath coupling, and evolution (II) is chosen at the intersection of the two blobs in the weak system-bath coupling regime.  }
    \label{fig:short-coh-de0}
\end{figure*}

Figure \ref{fig:short-coh-de0} shows the error in the coherences $\varepsilon_{coh}$ at short times.
We see that, as expected, gFT does not provide a good theory for the coherences for short times.
Nonetheless, the error remains contained at short times $\varepsilon_{coh}<0.08$.
Interestingly, the region with small error at short times does not agree with the expected strong system-bath coupling regime of applicability $\lambda >J$.
Indeed we observe that there is a region at small $\lambda$ where the error remains very small.
The evolution (II) illustrates this region, where we see exactly that, although at long times the coherence saturates to a large negative value.
This clearly cannot be captured by gFT due to the choice of the site basis.
However, for short times $t<t_c$ we see that the gFT evolution remains very close to HEOM.
It is also interesting to note that, in this region, HEOM predicts oscillatory behavior of the imaginary part, while gFT does not capture this feature. The phase diagram shows that there are two regions with larger error, corresponding to very large $\lambda$ and very large $J$. Evolution (I) illustrates the dynamics in one of these regions. We see that the gFT evolution starts deviating from HEOM already at $t<t_c$.
At long times the coherence is non-zero in HEOM, but, interestingly, it is smaller than that of evolution (II).
This shows that the short time validity of gFT cannot be inferred from the long-time results, and thus gFT at short times is not only valid for strong system bath coupling.
Indeed, the region with smaller error occurs at small $\lambda$.

\begin{figure*}[ht]
    \centering
    \includegraphics[width=\linewidth]{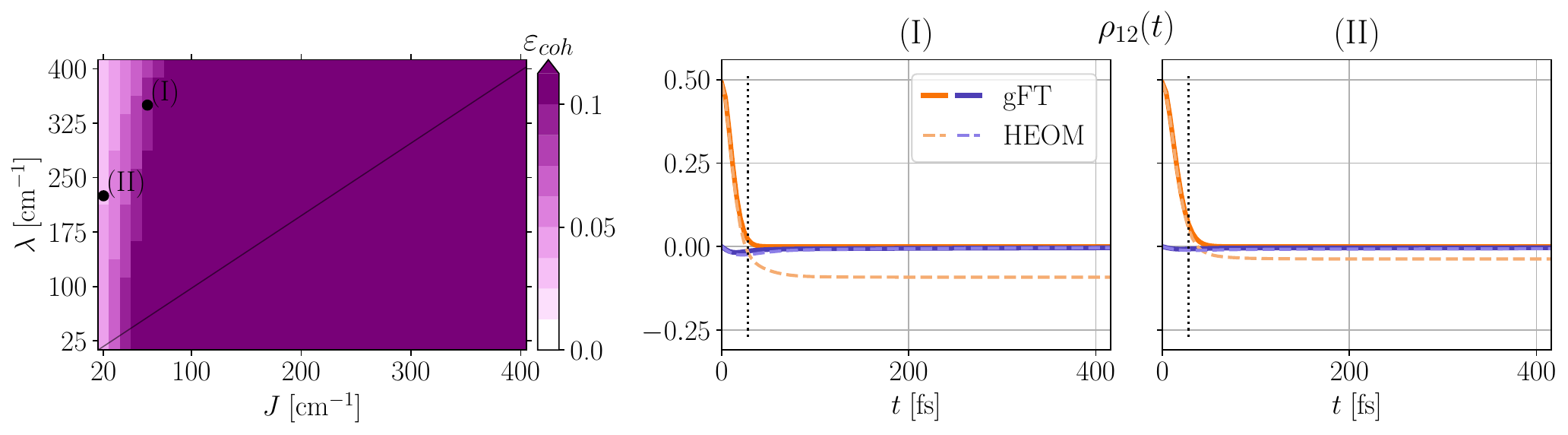}
    
    \caption{Phase diagram (left) for the intermediate time difference of coherences $\varepsilon_{coh}$ between generalized Förster Theory and HEOM for degenerate levels $\Delta E = 0\, \mr{cm}^{-1}$. Plots (I) and (II) depict two example time evolution comparing gFT (solid) and HEOM (dashed). Both of these evolutions are chosen at extremely weak hopping $J$, where the delocalization in the site basis is very small. They depict a case where gFT is still quite accurate (II) and a case with larger error $\varepsilon_{coh} \approx 0.1$ (I).}
    \label{fig:inter-coh-de0}
\end{figure*}

Figure \ref{fig:inter-coh-de0} shows the long-time error of the coherence. As expected and already discussed, gFT cannot model long time coherence in the site basis. For this reason, we see that the error is large $\varepsilon>0.1$ almost in the phase diagram. The only region where the error is $\varepsilon<0.1$ is for very small $J$, where the delocalization in the site basis will be very small.
An example of this is evolution (II) where we see that the coherence saturates to a non-zero, but small value. The example evolution (I) is close to $\varepsilon_{coh}=0.1$, and thus serves as an illustration for what a trace distance of $\varepsilon \approx 0.1$ physically represents, thus justifying our choice not to show these evolutions and focus on the ones where gFT is a good model.

\subsection{Intermediate energy difference $\Delta E =50\;{\rm cm}^{-1}$}

\begin{figure*}[ht]
    \centering
    \includegraphics[width=\linewidth]{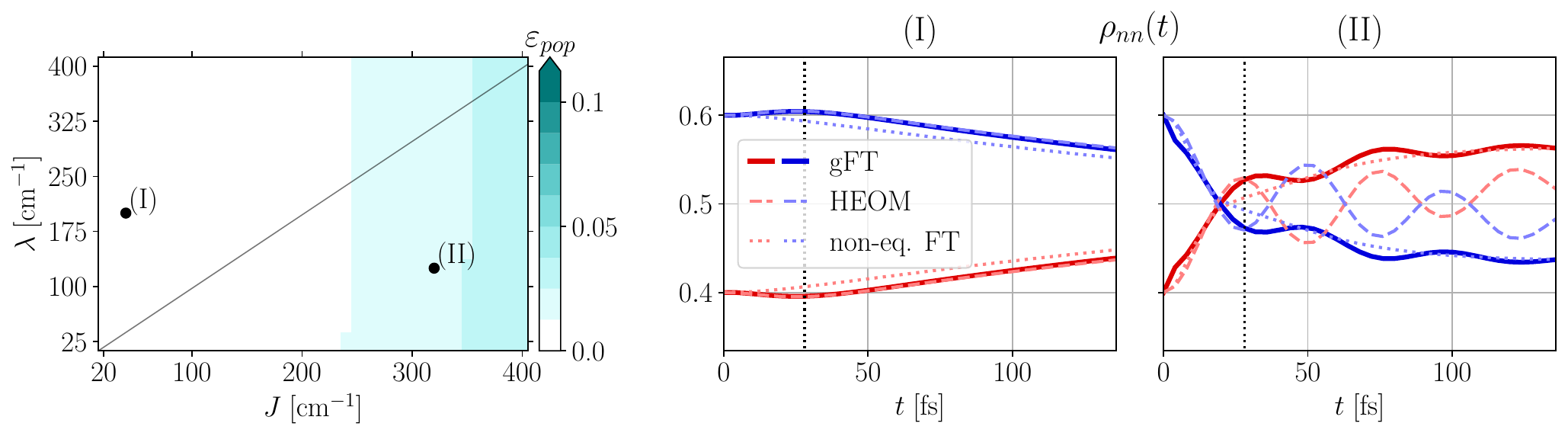}
    
    \caption{Phase diagram for the short time population error between gFT and HEOM (left) for $\Delta E = 50 \, \mr{cm}^{-1}$ (same as in Fig 1 of main text). Evolution (I) shows an example where the short time error is very small $\varepsilon_{pop}<0.0125$, note that non-eq FT already deviates from HEOM at short times. Evolution (II) shows a case at weak system-bath coupling, where the error is a bit larger, but gFT still remains relatively close to HEOM at short times.}
    \label{fig:short-pop-de50}
\end{figure*}

Figure \ref{fig:short-pop-de50} shows the error phase diagram for an intermediate energy difference between the two dimers of $\Delta E = 50$.
The behavior is fairly similar to the one observed in Fig. \ref{fig:short-pop-de0}, with a very small error at short times $\varepsilon_{pop}<0.04$.
The two example evolutions display the difference at short times, highlighting a very accurate evolution (I) with a very small error.
Although non-eq. FT also works well in this regime, we see that gFT is closer to HEOM.
Evolution (II) shows an evolution where the short time error is larger, and indeed we see that HEOM and gFT show a small deviations.

\begin{figure}[ht]
    \centering
    \includegraphics[width=\linewidth]{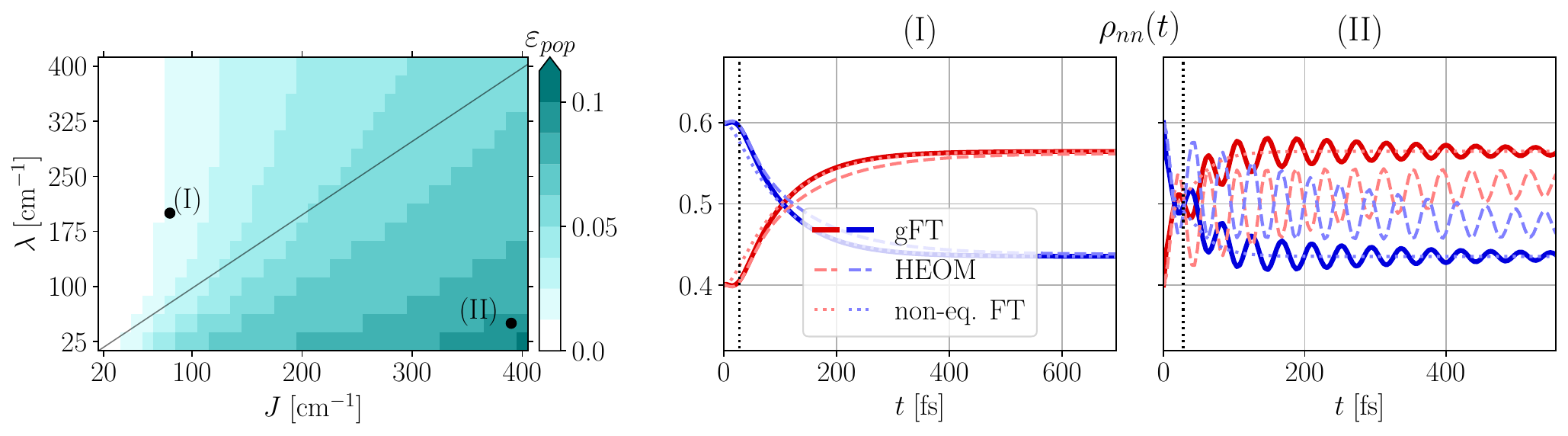}
    
    \caption{Phase diagram for the intermediate time population error $\varepsilon_{pop}$ (left) between gFT and HEOM for $\Delta E = 50 \mr{cm}^{-1}$ (same as in main text). Note that at intermediate times the validity of gFT aligns with the usual strong system-bath coupling regime. Evolution (I) depicts a case where the error is very small $0.0125<\varepsilon_{pop} <0.025$ and we see that this error happens around $t \approx 200 \mr{fs}$  where the population of HEOM grows more slowly towards the steady state. Evolution (II) shows a case with larger error $\varepsilon_{pop}\approx 0.1$ arising at very weak system-bath coupling. In this we see that gFT oscillates around a different value than HEOM since gFT converges to canonical equilibrium in the site basis.}
    \label{fig:inter-pop-de50}
\end{figure}

Figure \ref{fig:inter-pop-de50} shows the error of the population at intermediate times. This phase diagram clearly shows that gFT is valid for strong system bath coupling $\lambda>J$.
However, we  see that the boundaries of the regions with similar error do not seem to be simple linear functions of $J$.
This suggests that the gFT is valid when $\lambda > f(J)$, where $f(J)$ is a nonlinear function, which will generally depend on the rest of the parameters $\Delta E, \, k_{\rm B}T, \, t_c$.
We see however, that gFT is valid for quite a wide range of parameters. As stated in the main text, it should be noted that gFT has smaller error at \emph{intermediate} times than non-eq. FT at \emph{short} times.
Evolution (I) shows an example where gFT captures the HEOM dynamics well, with only small deviations for $100\; {\rm fs} \lesssim t \lesssim 300\; {\rm fs}$.
Non-eq. FT shows a deviation from HEOM and gFT at short times, but then agrees perfectly with gFT. The reason for this is simply that the parameters are in the regime of applicability of Förster $\lambda>J$. 
Evolution (II) shows a case where the error is larger $0.075<\varepsilon_{coh}<0.1$. In this case we see that gFT predicts some of the oscillatory behavior of HEOM. Although the frequency of the oscillations is reasonably well captured by gFT, the value around which they oscillate is not the same as in HEOM.
This could be caused by the fact that gFT always predicts an equilibrium state in the site basis, so the populations saturate to $1/(1 + e^{\pm \beta \Delta E})$, while HEOM can account for equilibrium in a different basis $\rho = e^{-\beta H'}/Z_\beta$. In comparison, we see that non-eq. FT does not predict oscillations, but saturates to the same equilibrium value as gFT.

\begin{figure}[ht]
    \centering
    \includegraphics[width=\linewidth]{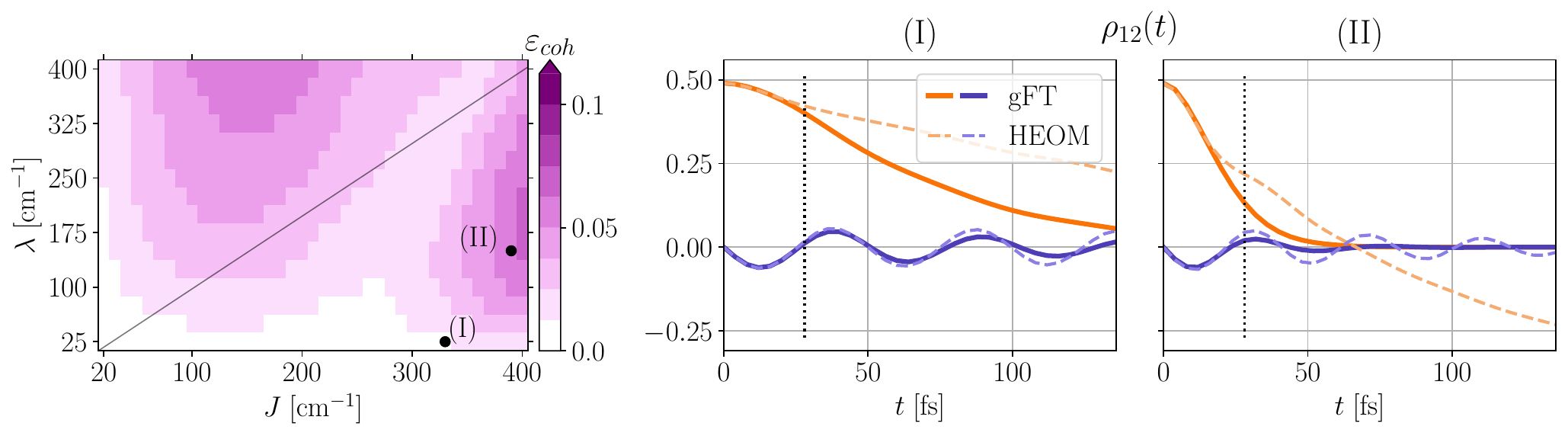}
    
    \caption{Phase diagram (left) for the short time difference of coherences $\varepsilon_{coh}$ between gFT and HEOM for $\Delta E = 50\, \mr{cm}^{-1}$. Plots (I) and (II) depict two example time evolution comparing gFT (solid) and HEOM (dashed). Evolution (I) is taken in the region of small error at weak system system bath coupling, where we see gFT and HEOM agreeing quite well until $t= t_c$. Evolution (II) is chosen in the blob of weak system-bath coupling, where we see a larger deviation of HEOM from gFT, specially on their real part.}
    \label{fig:short-coh-de50}
\end{figure}

Figure \ref{fig:short-coh-de50} shows the error in the coherences at short times for $\Delta E =50$. The phase diagram is quite similar to Fig. \ref{fig:short-coh-de0} showing two regions of high error for large $\lambda$ and for large $J$, and a low error region at small $\lambda$. Evolution (I) illustrates the behavior in this low error region, we see that the real part of the coherence for gFT closely follows HEOM until they depart right before $t_c$. Interestingly, the evolution of the imaginary part is accurate for longer, capturing the oscillations. This is not the case for (II) where we see a big departure of the HEOM coherence from the gFT, also for  the imaginary part.

\begin{figure}[ht]
    \centering
    \includegraphics[width=\linewidth]{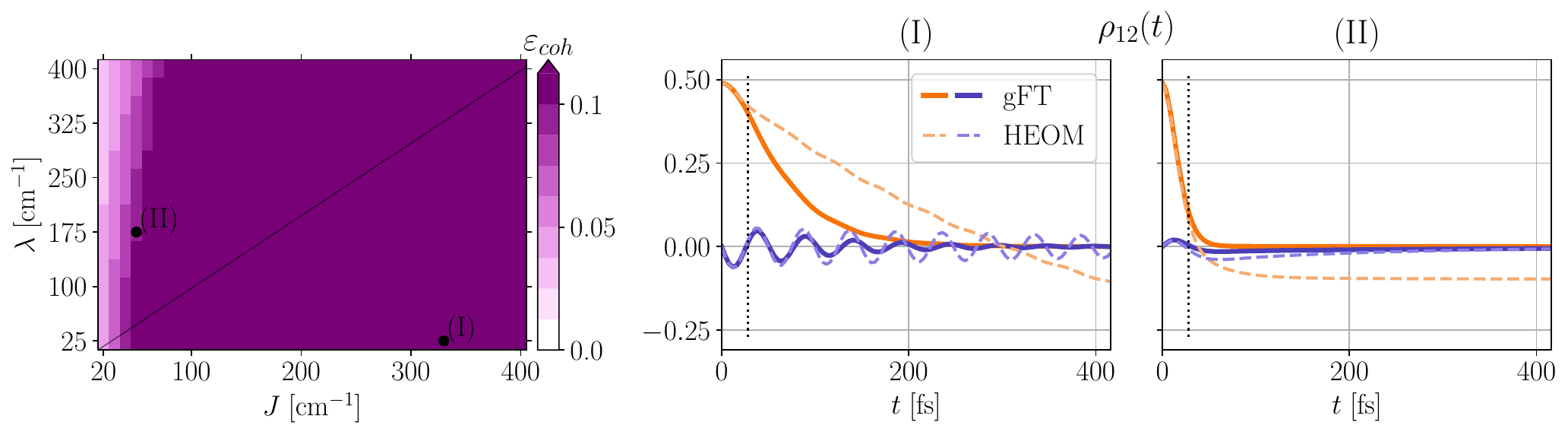}
    
    \caption{Phase diagram (left) for the intermediate time difference of coherences $\varepsilon_{coh}$ between gFT and HEOM for $\Delta E = 50\, \mr{cm}^{-1}$. Evolution (I) is the same as in Fig. \ref{fig:short-coh-de50}. Evolution (II) shows an error of $\varepsilon_{coh}\approx 0.1$.}
    \label{fig:inter-coh-de50}
\end{figure}

Figure \ref{fig:inter-coh-de50} shows that at intermediate times, the coherence dynamics predicted by gFT are completely different from those of HEOM, as we expected.
Evolution (I) is the same as in Fig. \ref{fig:short-coh-de50} and we see that HEOM shows long lived oscillations of the imaginary part, which die out in gFT, and the real part does not agree at all between the two cases. On the contrary, evolution (II), although it also has a considerable error, shows that this error comes mostly from the saturation value of the real part of the coherence due to delocalization. 

\subsection{Large energy difference $\Delta E =300\;{\rm cm}^{-1}$}
\begin{figure}[ht]
    \centering
    \includegraphics[width=\linewidth]{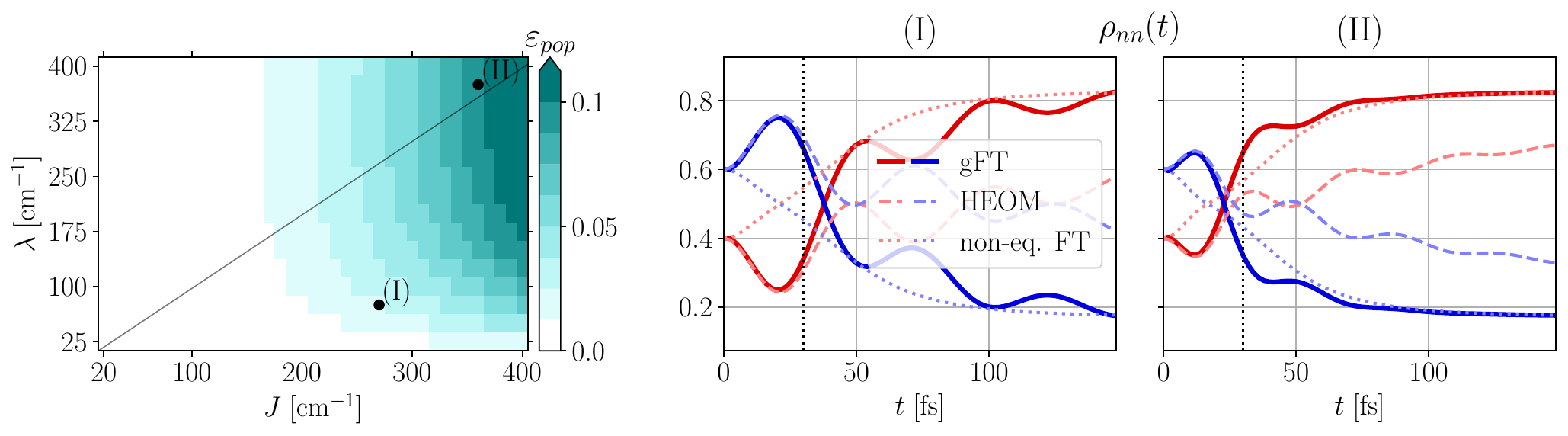}
    
    \caption{Phase diagram for the short time population error $\varepsilon_{pop}$ between gFT and HEOM (left) for $\Delta E = 300 \, \mr{cm}^{-1}$. Evolution (I) shows an example at weak system-bath coupling with very small error $\varepsilon_{pop}\approx 0.025$ with large error from non-eq FT. Evolution (II) shows a case with larger error $\varepsilon_{pop} \approx 0.1$ where gFT and HEOM deviate after the first bump.}
    \label{fig:short-pop-de300}
\end{figure}

Figure \ref{fig:short-pop-de300} shows the population error at short times for a large energy difference between the two energies of the dimer.
We observe that in this case, even the short time error can be large, with some values $\varepsilon_{pop}>0.1$ for very strong $J$ and strong $\lambda$.
The phase diagram also illustrates how the region with small $J<150$ has very small short time error and how the region with very small $\lambda\approx 25$ also has small error.
This confirms, once more, that the short time validity of gFT is not restricted to strong system-bath coupling.
Evolution (I) shows a case where the short time error is small $\varepsilon_{pop}$.
It shows a distinctive feature of the large $\Delta E$ case, the presence of a peak at short times, which arises due to the inhomogeneous term $\hat I(t)$ and thus cannot be captured by non-eq. FT. We therefore see that gFT captures the initial peak, describing much better the short time behavior of HEOM. At longer times, as we already saw before, gFT shows some oscillations, which are also shown by HEOM, but which are not at the same values of the population. In this case non-eq. FT performs very poorly and does not approximate the HEOM evolution at any time. Evolution (II) is an example that even modelling the initial peak well does not guarantee that the short time evolution will be well described, since we see the two evolutions separating already before $t_c$. 

\begin{figure}[ht!]
    \centering
    \includegraphics[width=\linewidth]{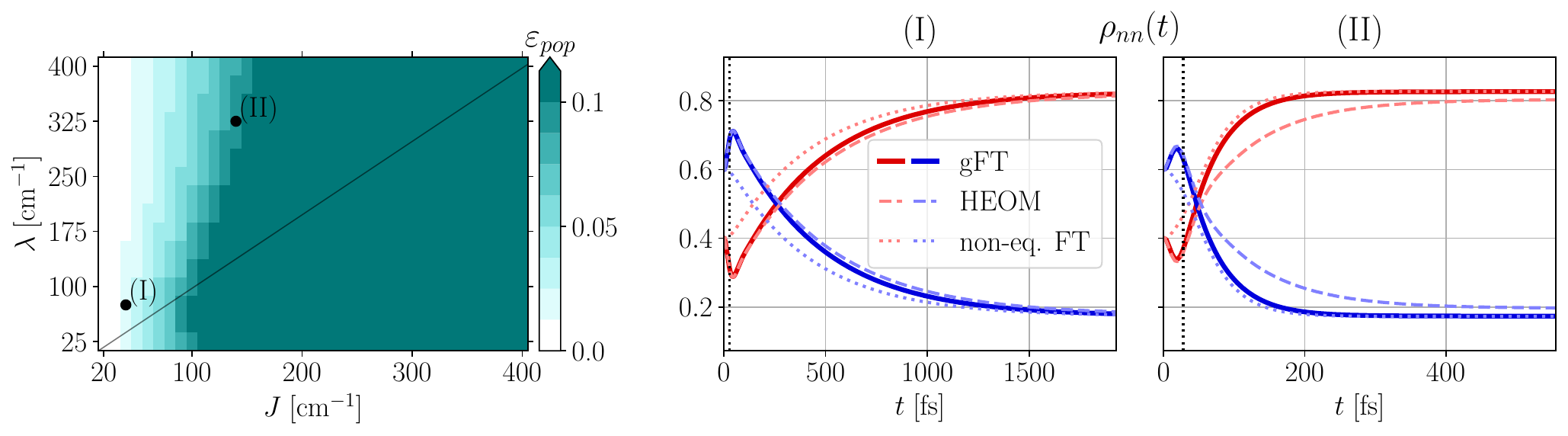}
    
    \caption{Phase diagram for the intermediate time population error $\varepsilon_{pop}$ between gFT and HEOM (left) for $\Delta E = 300 \, \mr{cm}^{-1}$. Evolution (I) shows a case with low error, which non-eq FT fails to model accurately. Evolution (II) shows a case with larger error $\varepsilon_{pop} \approx 0.1$ where again HEOM shows a slower relaxation towards equilibrium than Förster theories.}
    \label{fig:inter-pop-de300}
\end{figure}

Figure \ref{fig:inter-pop-de300} shows the population error at intermediate times for large $\Delta E$. We see that in this limit, a large part of the phase diagram has large error $\varepsilon_{pop}>0.1$. This suggests that the validity of gFT for intermediate times critically depends on $\Delta E$. As expected the regime of validity of the theory is for strong system bath coupling $\lambda>J$. Evolution (I) shows an example where the presence of the short time peak separates the predictions of gFT and non-eq. FT so that only gFT provides a good model for the HEOM evolution. However, in evolution (II) we see an example where HEOM deviates from gFT after the populations cross.  

\begin{figure}[ht!]
    \centering
    \includegraphics[width=\linewidth]{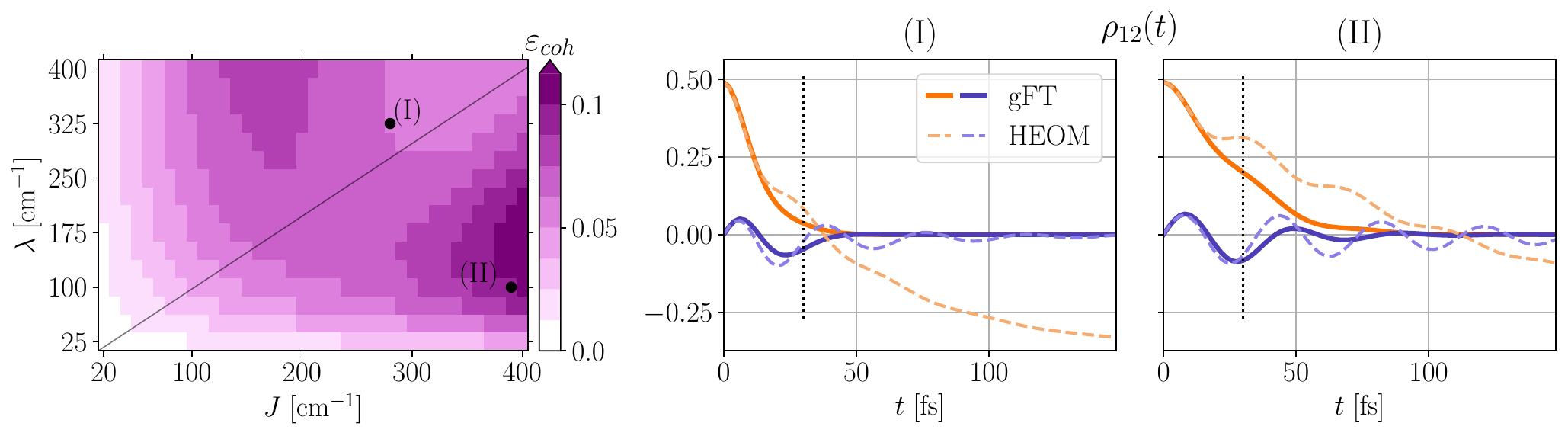}
    
    \caption{Phase diagram (left) for the short time difference of coherences $\varepsilon_{coh}$ between gFT and HEOM for $\Delta E = 300\, \mr{cm}^{-1}$. Evolution (I) shows an example with moderate error in between the two blobs, and evolution (II) shows an example with large error in the weak system-bath coupling blob. We again see that most of the error comes from the real part of the coherence.}
    \label{fig:short-coh-de300}
\end{figure}

Figure \ref{fig:short-coh-de300} shows that for large $\Delta E$, the region where the short time coherence is well captured moves to the lower left corner of the phase diagram with small $\lambda$ and $J$.
The phase diagram still retains two lobes with a high error, although now the large $J$ has a higher error value value. Evolution (I) illustrates the deviations at short times for large $J$ and large $\lambda$, while evolution (II) shows that the real part of the coherence acquires some oscillatory behavior, which therefore gives a larger error at short times. 

\begin{figure}[ht!]
    \centering
   \includegraphics[width=\linewidth]{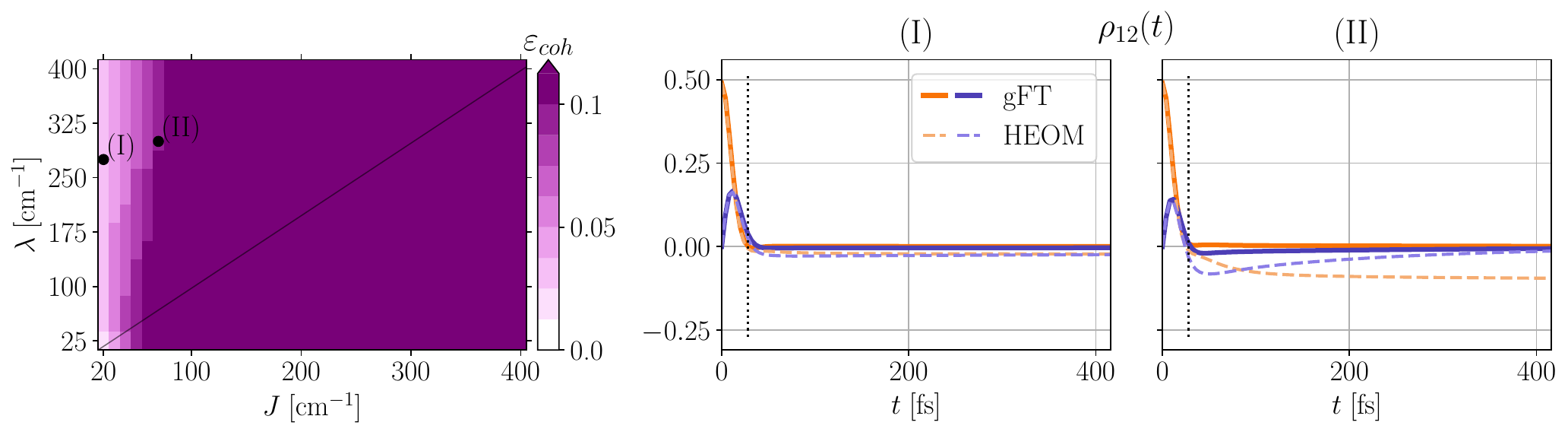}
    
    \caption{Phase diagram (left) for the intermediate time difference of coherences $\varepsilon_{coh}$ between gFT and HEOM for $\Delta E = 300\, \mr{cm}^{-1}$. Evolution (I) shows an example at very small $J$ where the error is small, and evolution (II) an example with larger error $\varepsilon_{pop}\approx 0.1$.}
    \label{fig:inter-coh-de300}
\end{figure}

Lastly, Fig \ref{fig:inter-coh-de300} shows a similar behavior for the intermediate time coherence error to the other values of $\Delta E$. Only systems with very small $J$ have a relatively small intermediate time coherence error. Evolutions (I) and (II) represent two examples of the low error region and at the boundary, respectively. 

\end{document}